\documentclass{appolb}
\usepackage{graphicx}
\usepackage{amsmath,amssymb}
\usepackage{slashed}
\usepackage{bm}
\usepackage[colorlinks,linkcolor=blue,citecolor=blue]{hyperref}

\newcommand{\be}{\begin{equation}}
\newcommand{\ee}{\end{equation}}
\newcommand{\bea}{\begin{eqnarray}}
\newcommand{\eea}{\end{eqnarray}}
\newcommand{\ba}{\begin{eqnarray}}
\newcommand{\ea}{\end{eqnarray}}

\begin{document}
\title{Instanton effects in  Euclidean vacuum,\\  real time production  and in the light front wave functions}

\thanks{Presented at 2021 Zakopane school}%

\author{Edward Shuryak, Ismail Zahed
\address{Center for Nuclear Theory, Stony Brook University \\
Stony Brook NY 11794 USA}
}
\maketitle
\begin{abstract}
Nontrivial topological structures of non-Abelian gauge fields were discovered in the 1970's. Instanton solutions,
describing vacuum tunneling through topological barriers, have fermionic zero modes which are at the origin of 
$^\prime$t Hooft effective Lagrangian. In the 1980's instanton ensembles have  been used to explain chiral symmetry breaking.
In  the 1990's a large set of numerical simulations were performed deriving Euclidean correlation functions.
The special role of scalar diquarks in nucleons,  and color superconductivity in dense quark matter have been elucidated.

In these lectures, we discuss further developments of physics related to gauge topology. 
We show that the instanton-antiinstanton ``streamline" configurations describe ``sphaleron transitions" in high energy collisions, which
result in production of hadronic clusters with nontrivial topological/chiral charges. (They are not yet observed, but
discussions of dedicated experiments at LHC and RHIC are ongoing.)  

Another new direction is instanton effects in hadronic spectroscopy, both in the rest frame and on the light front.
We will discuss their role in central and spin-dependent potentials, formfactors and  antiquark nuclear ``sea".
Finally, we summarize the advances in the semiclassical theory of deconfinement,
 and chiral phase transitions at finite temperature,
in QCD and in some of its ``deformed" versions.

\end{abstract}
\PACS{PACS numbers come here}

\section{Brief overview of gauge topology}
\begin{figure}[htbp]
\begin{center}
\includegraphics[width=14cm]{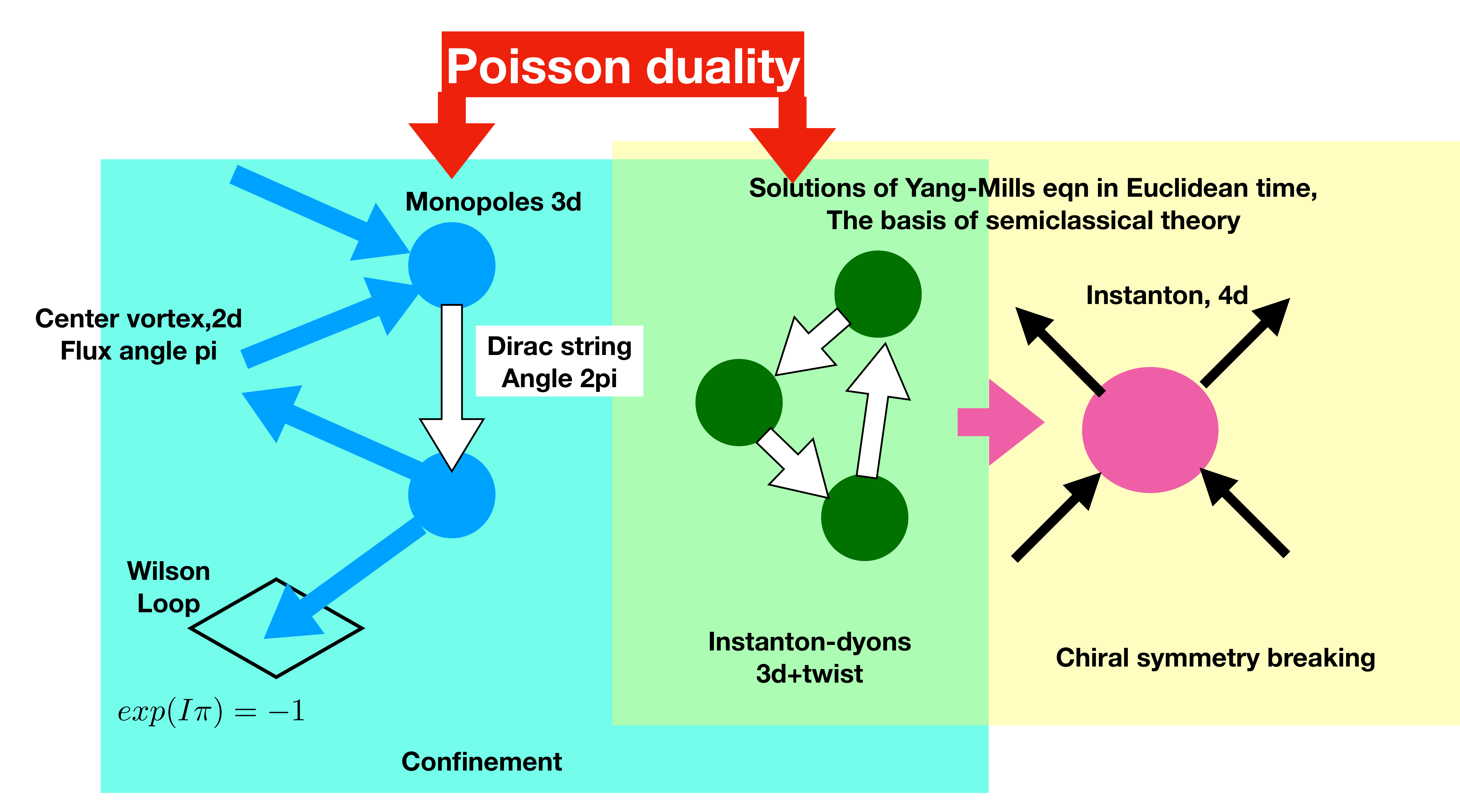}
\caption{Map of gauge topology, explanations are in the text.}
\label{fig_topo_map}
\end{center}
\end{figure}

We  start these lectures with a map, outlining the role and interdependence of various
topological structures in the QCD vacuum.
The phenomenon of {\em color confinement}  (left blue region in Fig.\ref{fig_topo_map}) 
was first related with {\em center vortices}, associated with a phase $\pi$
for a quark looping around, changing the sign of the linking Wilson line. Two such vortices combined together,
 lead to a singularity with a phase
$2\pi$, shown by white an arrow, known as  the {\em Dirac string}. Their ends are
identified as {\em magnetic monopoles} (blue disks). Confinement is 
 their {\em Bose-Einstein condensation}, perhaps the most physical 
signature of this phenomenon. 

The right yellow region of Fig.\ref{fig_topo_map} is associated with another major
nonperturbative phenomenon, {\em chiral symmetry breaking}.  The pink disk refers to
{\em instantons}, 4D solitons in Euclidean space-time. Fermions in this field have 
{\em zero modes}, elevating each instanton into  $^\prime$t Hooft multi-fermion operator.
The four black arrows correspond to the case of two quark flavors, $u,d$. The resulting
4-fermion vertex is similar to Nambu-Jona-Lasinio hypothetical interaction, and also
breaks spontanously $SU(N_f)_a$ chiral symmetry provided the instanton density
is sufficiently large. 

At finite temperature the {\em Polyakov line} has a number of eigenvalues 
or $holonomies$ $\mu_i(T)$ (see below). The instanton solution amended by such asymptotics for the $A_0$ field,  splits into instanton constituents, 
known as {\em instanton-dyons} or {\em instanton-monopoles}\footnote{Both names were criticized: they are neither the dyons of  Schwinger, nor monopoles of Dirac, but
Euclidean self-dual objects, with equal  $E$ and $B$ up to a sign. Perhaps a new name, emphasizing their Euclidean nature, is needed: can it e.g. be $instantino$? } . 
Like instantons, they have topological charges. Unlike instantons, the charges  are
{\em not quantized to integers} $Q$. This is possible since they are connected by Dirac
strings due to their magnetic charges.
Ensembles of these topological objects will be shown to generate both confinement, and chiral breaking
at the end of these lectures. 

Last but not least is the phenomenon of {\em Poisson duality}, started as a technical observation
relating  partition functions derived in terms of monopoles and  in terms of instanton-dyons. 
As we will discuss below, it has wider meaning:
their equality  brings to mind equivalence of two other great
approaches to classical dynamics, by  Hamilton and Lagrange, respectively.

The detailed discussion of these configurations, in addition to discussions about  the QCD flux tubes and
holographic QCD etc, can be found in~\cite{Shuryak:2021vnj}.

\section{Brief history of instantons}
\label{intro}
\subsection{Semiclassical theory based on path integrals: instantons and fluctons} 
The quantum mechanics courses include semiclassical methods based on certain
representation of the wave function, starting with the celebrated Bohr-Sommerfeld 
quantization condition, applied to the oscillator and hydrogen atom, and followed by
 Wentzel-Kramers-Brillouin
 (WKB) approximation, developed in 1926.  Unfortunately, subsequent study has shown that
 generalization of those to physical systems with more than one degree of freedom, as well as to
 systematic order-by-order account for quantum fluctuations,  are not possible.
 
 By definition the Feynman path integral gives the {\em density matrix} in coordinate representation
(see e.g. a very pedagogical  Feynman-Hibbs book) 
\be \label{denmat}
\rho(x_i,x_f,t_{tot})\ =\ \int_{x(0)=x_i}^{x(t_{tot})=x_f} Dx(t) e^{i\,S[x(t)]/\hbar} \ .
\ee
Note that it is a function of the initial and final coordinates, as well as the time needed for the transition. Here $S$ is the classical action of the system, e.g. for a particle of mass $m$ in a static potential $V(x)$ it is
\[  S \ = \ \int_0^{t_{tot}}dt\, \bigg[ \frac{m}{2}{\bigg(\frac{dx}{dt}\bigg)}^2 - V(x)    \bigg] \ ,\]
Feynman has shown that the oscillating exponent along the path, provides the correct weight of the paths integral (\ref{denmat}). For reference, the same object can also be written in a form closer to that used in quantum mechanics courses. Heisenberg  wrote it as a matrix element of the time evolution operator, the exponential of the Hamiltonian~\footnote{Here, we assume that the motion happens in a time-independent potential, for otherwise the exponential is time ordered.}
\be \rho(x_i,x_f,t_{tot})\ =\langle x_f | e^{i\hat H t_{tot}} | x_i \rangle \ee
between states in which a particle is localized  in-out.

Schroedinger set of stationary states $\hat H  | n \rangle = E_n | n \rangle$ can also be used
as a basis set. Because the Hamiltonian is diagonal in this basis, there is a single (not double)
sum 
  \be \rho(x_i,x_f,t)\ =\sum_n \psi_n^*(x_f)  \psi_n(x_i ) e^{i E_n t} \label{psipsi}\ee
with $\psi_n(x)= \langle n | x \rangle$. 
Oscillating weights for different states are often hard to calculate, and one may
wonder if it is possible to analytically  continue in time,  to the  Euclidean version
with $i$ absorbed into it. For reasons which will soon become clear, we will
also define this imaginary time on a circle with circumference $\beta$
 $$\tau=i\,t\in [0,\beta]$$
In this way we will be able to describe
 $quantum+statistical$ mechanics of a particle in a heat bath with temperature $T$
 related to the
circle circumference $$\beta={\hbar \over T}$$
  Such periodic time is known as the Matsubara time. 
The expression (\ref{psipsi})  is now
    \be \rho(x,x,t)\ =\sum_n |\psi_n(x)|^2 e^{- E_n /T} \ee
    which sums the  quantum-mechanical probabilities to find a particle at point $x$, times  its   thermal weight.  Performing the integral over all $x$,
     and using the normalization of the weight functions, 
    one finds the expression for the thermal partition function 
    \be Z=\sum_n e^{- E_n /T} \ee
   We will use this expression below  in a  Feynman path integral form, with
   (i) taken over all $periodic$ paths with the same endpoints, and with (ii) Euclidean or rotated time.
   The probability to find a  particle at a certain point is then  
\be
P(x_0,t_{tot}) =\int_{x(0)=x_0}^{x(\beta)=x_0} Dx(\tau) e^{-S_E[x(\tau)]/\hbar} \ .
\label{P}
\ee
Note that here, the exponent is not oscillating and equals the  Euclidean action
 \be S_E \ = \ \int_0^{\beta}d\tau\, [ \frac{m}{2}{(\frac{dx}{d\tau})}^2 + V(x)] \ee
 in which the sign of the potential is reversed, and the time derivative is understood to be over $\tau$.
 
 Looking for the periodic paths
with the minimal action, one may start with the simplest, i.e.  a
particle at rest with  $x(\tau)=const$! Such path is dominant 
for small time (Matsubara circle), with $\beta\rightarrow 0$ (or high temperature $T$)\footnote{ Note that it is
opposite to the limit discussed above in the Hamiltonian approach.}
 If one  ignores the time dependence and velocity on
 the paths, there is no kinetic term and only the potential one in the action contributes. So
\be
P(x_0,\beta) \sim e^{- \frac{V(x_0)}{T}}\ ,
\ee
which corresponds to classical\footnote{Note that if we would keep $\hbar\neq 1$, the one in
$\beta$ and in the exponent $exp(-S/\hbar)$ would cancel out, confirming the classical nature of this limit. 
} thermal distribution for a particle in a potential $V$.

\begin{figure}[b]
\begin{center}
\includegraphics[width=10cm]{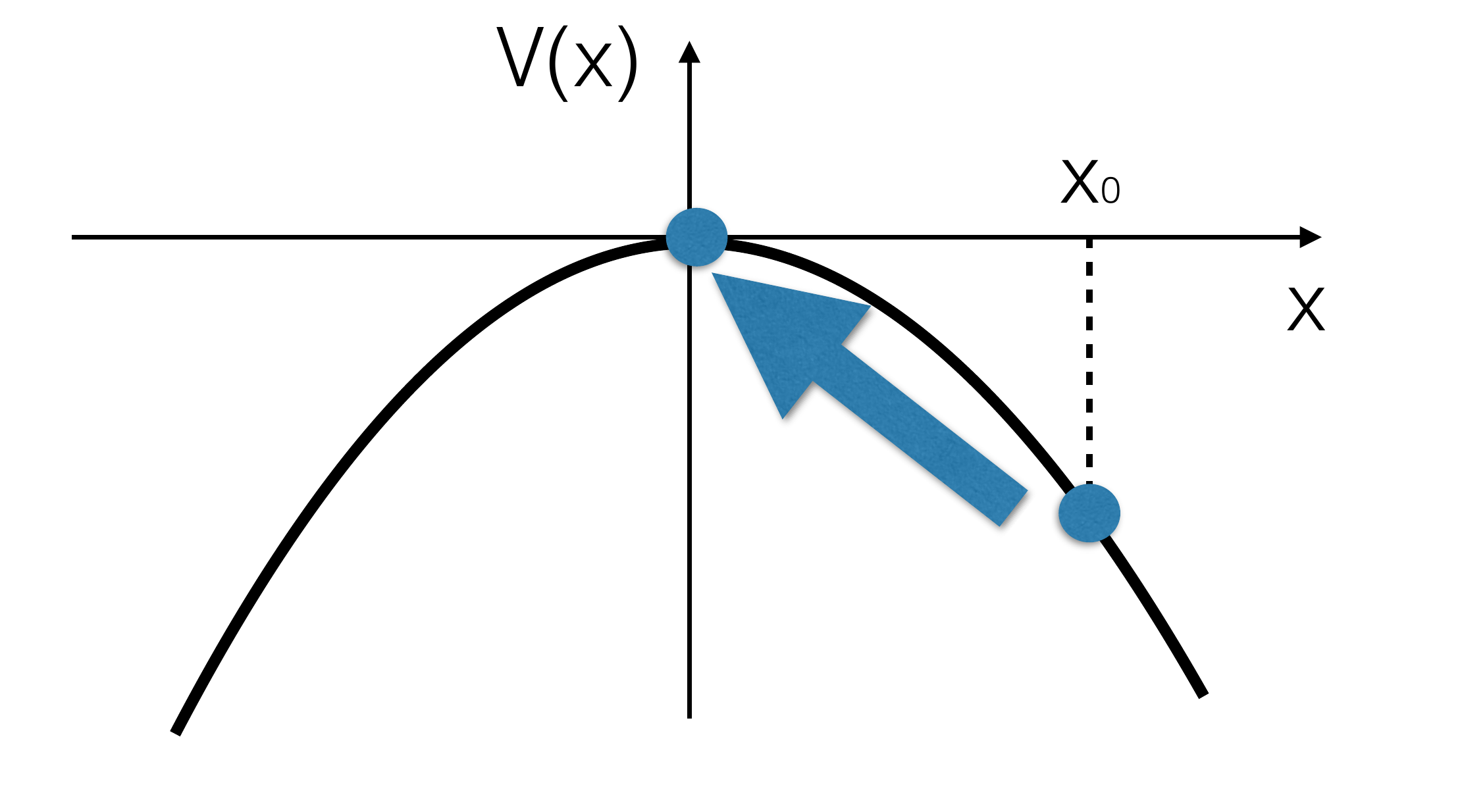}
\caption{Sketch of the flucton path climbing toward the (flipped)  minimum of the potential.}
\label{fig_fluct}
\end{center}
\end{figure}

 Since the weight is
$exp\big(-S[x(\tau)]\big)$, the paths with the smallest action should give the largest contribution.  
These paths satisfy the classical (Euclidean) equation of motion, as we will carry below. The semiclassical approximation  -- the $dominance$ 
 of these paths -- would be justified, as soon as the corresponding action is large
\be S_{cl} \equiv S[ x_{cl}(\tau)] \gg \hbar \ee
Such classical paths were called {\em fluctons} in \cite{Shuryak:1987tr}. 
These paths should have Euclidean time period $\beta=\hbar/T$. For simplicity, 
let us start with ``cold" QM, or vanishingly small $T$,  $\beta\rightarrow \infty$.
Little thinking of how to arrange a classical path with a very long period 
leads to the following solution: the particle should roll to the top of the 
(flipped) potential with exactly such energy as to sit there for very long time, before
it will roll back to the  (arbitrary) point $x_0$ from which the path started. 
The classical paths corresponding to relaxation toward the potential bottom take the form of a path
{\em ``climbing up"}  from an arbitrary point $x_0$ to the  maximum, see Fig. \ref{fig_fluct}

Let us start with the harmonic oscillator, as the unavoidable first example.
For simplicity, let us use the canonical units, 
in which the particle  mass is $m=1$,  and the
oscillator frequency is $\Omega=1$, so that our (Euclidean) Lagrangian
\footnote{Note again the flipped sign of  the potential term. In Minkowski time the potential has sign minus.}
 is simplified to
\be
  L_E = \frac{\dot{x}(\tau)^2}{2} +  \frac{x(\tau)^2}{2} \ .
\ee
Because of this sign, in Euclidean time $\tau$ the oscillator does not oscillate $e^{it}$ but relaxes $e^{-\tau}$.  For a harmonic oscillator, the classical equations
 of motion (EOM) are of course not difficult to solve. However, 
 it is always easier to get solutions using energy conservation (first integral of motion for static potentials).
 Since we are interested in solutions with zero energy $E=0$,
  they correspond to $\dot{x}^2=2V(x)$. The boundary conditions are $x_0$, 
 at $\tau=0$ plus periodicity
on a circle with circumference $\beta$. This solution is
\be
   x_{flucton}(\tau)\ =\ x_0\ \frac{\left( e^{\beta - \tau}\ +\ e^\tau \right)}{e^\beta + 1}
    \ .
\ee
defined for $\tau \in [0,\beta]$. The particle moves toward $x=0$ and reaches some minimal value, at  $\tau=\beta/2$, and then returns to the initial point $x_0$ again at $\tau=\beta$. 
 Due to the periodicity in $\tau$, one may shift its range to  $\tau \in [-\beta/2,\beta/2]$. 
The minimal value at $\tau=\beta/2$ $$ x_{min}={x_0 \over cosh(\beta/2)} \rightarrow_{\beta\rightarrow\infty} 0 $$
is exponentially small at low temperature. The climbing to the potential top at $x=0$ is nearly accomplished, if the period  is large.

The solution in the zero temperature or   $\beta\rightarrow \infty$ limit
simplifies to  $x_0 e^{-| \tau |}$.
In the opposite limit of small $\beta$ or high $T$, there is no time to move far from $x_0$,
so in this case the particle does not move at all.
The  classical action of the flucton path is
\be
   S_{flucton}\ =\ x_0^2 \ \tanh\bigg(\frac{\beta}{2}\bigg) \ ,
\ee
The result means that  the particle distribution
\be
  P(x_0) \sim \exp\left(- \frac{x_0^2}{\coth (\frac{\beta}{2})}\right) \ ,
\ee
is Gaussian at any temperature. Note that the width of the distribution
\be
 <x^2>\ =\ \frac{1}{2} \coth \bigg(\frac{\beta}{2}\bigg)\ =
     \ \frac{1}{2} + \frac{1}{e^\beta -1}\ ,
\ee
is  the ground state energy plus the contribution due to the thermal excitations, which
we already mentioned at the beginning of the chapter. So, the flucton reproduces
the well known results for the harmonic oscillator, see e.g. Feynman's Statistical Mechanics. 

Flucton solutions for several well used QM problems (anharmonic oscillator,  the quartic double well, sin-like potential) were discussed in detail in \cite{Escobar-Ruiz:2016aqv}. Unlike WKB, one can define and calculate 
semiclassical series in $\hbar$ by well-defined procedure, using Feynman diagrams. Finite temperature
fluctons were used for anharmonic oscillator and 4-nucleon clustering in \cite{Shuryak:2019ikv}.

Quantum mechanical instantons were introduced\footnote{It is interesting that the first application to the QFT
problem -- pairs production in constant electric field -- were done already in 1931, by Sauter \cite{Sauter:1931zz} . }
  by \cite{Polyakov:1976fu} in the context of tunneling in the double-well potential. In the inverted
  potential it is a path going from one maximum to another.
  
  For an early pedagogical review, including the one-loop (determinant) 
calculation 
by the same tedious scattering-phase method see ABC's of instantons ~\cite{Vainshtein:1981wh}. Two-loop 
corrections were first calculated in~\cite{Aleinikov:1987wx},  some technical errors
in it were  corrected by Shuryak and Wohler  in
\cite{Wohler:1994pg}.  Three loop corrections have been calculated by 
Escobar-Ruiz, Turbiner and Shuryak in
\cite{Escobar-Ruiz:2015nsa}. 

The calculation goes in the same standard way, the path is written as a classical plus {\em quantum fluctuation}
$\delta x(\tau)$, and then the action is expanded in powers of $\delta x(\tau)$.
One technical (but important) difference between the $instantons$ and $fluctons$, is that in the former case
the {\em fluctuation operator} ($O(\delta x(\tau)^2)$)
has a {\em zero mode}, related to time-shift symmetry.  Therefore the Green function for the fluctuations, 
needs to be defined in a {\em nonzero mode subspace} of Hilbert space.  This complication leads to
new features, such as additional diagrams not following directly from the Lagrangian.

The Gauge theory instanton was found in  the famous work
by Belavin, Polyakov, Schwartz and Tyupkin, and since  known as the BPST instanton \cite{Belavin:1975fg}.
Let us recall how it was obtained, as some of the expressions will be useful for later use.
 To find the classical solution corresponding to tunneling, BPST  used  the following 4-dimensional spherical ansatz  depending on the
 $radial$ trial function $f$
\be g A_\mu^a=\eta_{a\mu\nu} \partial_\nu F(y), \,\,\,\, F(y)=2\int_0^{\xi(y)} d\xi'   f(\xi')     \ee
with $\xi=  ln(x^2/\rho^2)$ and $\eta$ the 't Hooft symbol
defined by 
\ba 
\label{eta_def}
\eta_{a\mu\nu}&=&\left\{ \begin{array}{rcl}
 \epsilon_{a\mu\nu} &\hspace{0.5cm}& \mu,\,\nu=1,\,2,\,3, \\
 \delta_{a\mu}      &              & \nu=4,  \\
-\delta_{a\nu}      &              & \mu=4.
\end{array}\right.
\ea
We also define $\overline\eta_{a\mu\nu}$ by changing the sign of
the last two equations. 
Upon substitution of the gauge fields in  the gauge Lagrangian $(G_{\mu\nu})^2$ 
one finds that the effective Lagrangian has the form
\be L_{eff}=    \int d\xi \left[{\dot{f}^2\over 2}+2f^2(1-f)^2 \right]
\ee   
corresponding to the motion of a particle in a double-well potential. The Euclidean solution is that of  a quantum mechanical instanton,
connecting the $maxima$ of the flipped potential.  The corresponding field is 
\be  
\label{BPST_inst} 
A^a_\mu(x)= {2\over g}{\eta_{a\mu\nu}x_\nu \over x^2+\rho^2}
\ee 
Here $\rho$ is an arbitrary parameter characterizing the size of
the instanton. Like in the potential we discussed in the preceding
section, its appearance is dictated by the scale invariance of  the classical
Yang-Mills
equation. The ansatz itself perhaps needs some explanation. The 't Hooft symbol projects to sefl-dual fields,
which best captured by the identity
\ba 
\label{Bog_ineq}
S &=& \frac{1}{4g^2} \int d^4x\, G^a_{\mu\nu} G^a_{\mu\nu} \;=\;
 \frac{1}{4g^2}\int d^4x\, \left[\pm G^a_{\mu\nu} \tilde G^a_{\mu\nu}
 + \frac{1}{2} \left( G^a_{\mu\nu}\mp \tilde G^a_{\mu\nu}\right)^2
  \right],
\ea
where $\tilde G_{\mu\nu}=1/2\epsilon_{\mu\nu\rho\sigma}G_{\rho\sigma}$
is the dual field strength tensor (the field tensor in which the roles
of electric and magnetic fields are interchanged). Since the first term
is a topological invariant (see below) and the last term is always 
positive, it is clear that the action is minimal if the field 
is (anti){\em  self-dual}\footnote{This condition is written in Euclidean
space. In Minkowski space there is an extra $i$ in  the electric field.}
\be 
\label{self_dual}
G^a_{\mu \nu}=\pm\tilde G^a_{\mu \nu},
\ee
In a simpler language, it means that the Euclidean electric and magnetic
fields are the same\footnote{In the BPST paper the self-duality condition (1-st order differential equation)
was solved, rather than (2-nd order) EOM for the quartic oscillator. }. The action density is given by
\be 
\label{g2_inst}
(G^a_{\mu\nu})^2 =\frac{192\rho^4}{(x^2+\rho^2)^4} .
\ee 
It is  spherically symmetric, and very well localized, 
at large distances it is $\sim x^{-8}$.
The action depends on the scale only via the running coupling
\be  S={8\pi^2 \over g^2(\rho) }
\ee
The one-loop (determinant) quantum corrections were calculated in the
classic paper by $^\prime$t Hooft \cite{tHooft:1976snw}. The two-loop and higher order quantum corrections have not been calculated to this day,
due to the difficulties with the part of the Green functions related to the  nonzero modes.

\subsection{Chiral symmetry breaking in QCD}
Chiral symmetries are additional symmetries  that  appear when quarks are massless.
Since the mass term is the only one in the QCD Lagrangian connecting spinors with right (R) and left (L)
chiralities, the $SU(N_f)$ symmetry of flavor quark rotations,  gets doubled by acting on the  L- and R-quarks 
separately.  Also, it can be  viewed as $vector$ and $axial-vector$ transformations, in which rotations
on    the L- and R-spinors in the same or opposite directions. The latters are further split into
$U(1)_A$ (acting on all quarks by $exp(i\gamma_5 \theta)$ and  $SU(N_f)_a$.

The physics of the nonperturbative vacuum of strong interaction started even before QCD.
 Nambu and Jona-Lasinio (NJL) \cite{Nambu:1961tp}, inspired by BCS theory of superconductivity, have  qualitatively explained that
 strong enough attraction of quarks can break  $SU(N_f)_a$ chiral symmetry spontaneously and, among many other effects, create
  near-massless pions.
  
Instantons, the basis for a semiclassical theory of the QCD vacuum and hadrons, has been discovered in 1970's
\cite{Belavin:1975fg}, and soon $^\prime$t Hooft has found their  fermionic zero modes and formulated his famous
effective Lagrangian \cite{tHooft:1976snw}. Not only it solved the famous ``$U_A(1)$ problem" -- by making the $\eta'$ non-Goldstone and heavy --   but it also produces
a strong attraction in the $\sigma$ and $\pi$ channels.
In the framework of the so called  instanton liquid model (ILM)~\cite{Shuryak:1982hk},
it provided  a microscopic (QCD-based) 
  basis for  chiral symmetry breaking,  chiral perturbation theory and the pion properties. Its two parameters
  \begin{equation} \label{eqn_n_rho}
  	 \rho=\frac 13 \,{\rm  fm},\qquad n_{I+\bar{I}}=\frac 1{R^4}=1 \, {\rm fm}^{-4}
 \end{equation} 
play the same role as  the cutoff  and coupling in the NJL model. These parameters 
 have withstood the scrutiny of time, and describe
rather  well the chiral dynamics related to pions, the Euclidean correlation functions in the few femtometers range, interacting ensemble of instantons,
and much more, see~\cite{Schafer:1996wv} for a review.

The subject of these lectures is many other uses of instantons discussed in recent works. 
We will consider quark pair production,  and its  role in the isospin asymmetry of the nucleon ``sea",
the inclusion of $\bar I I$ molecules in mesonic formfactors,  and forces between quarks in hadrons.

\subsection{Euclidean correlation functions}
 This analysis originally started from the small-distance
OPE and the QCD sum rule method, it moved to intermediate distances (see review
e.g. in  \cite{Shuryak:1993kg}),

First, the experimentally known correlation functions 
were reproduced by the instanton liquid model at {\em small distances} at a quantitative level. 
Then, for many mesonic channels~\cite{Shuryak:1992jz,Shuryak:1992ke}, significant
numerical efforts were made, allowing to calculate the 
relevant correlation functions  till larger distances (of about 1.5 fm),
where they decay by
few decades. As a result, the predictive power of the model has been
explored in substantial depth. Many of 
the coupling constants and even hadronic masses
were calculated, with good
 agreement with experiment and lattice.

\subsection{Diquarks and color superconductivity}

Subsequent calculations of baryonic correlators \cite{Schafer:1993ra}
has revealed further surprising facts. 
In the instanton vacuum the nucleon was shown to be made 
of a ``constituent quark" plus a deeply bound $diquark$, with a mass nearly the same
as that of constituent quarks. On the other hand, decuplet baryons (like
$\Delta^{++}$) had shown no such diquarks, remaining  weakly bound
set of three  constituent quarks. 
Deepling bound scalar diquarks are a direct consequence of the 
 't Hooft Lagrangian, a  mechanism  that is also shared by the Nambu-Jona-Lasinio
 interaction~\cite{Thorsson:1989fw}, but ignored for a long time.
 
It also leads to the realization that diquarks can become Cooper pairs in dense quark matter,
see \cite{Schafer:2000et} for a review on ``color superconductivity".

\section{The topological landscape and sphaleron production processes}
 There has been significant development in the understanding of the ``topological landscape" in 
relation between instantons and sphaleron production processes, see  \cite{Ostrovsky:2002cg}.
Recently there have been renewed interest, due to the possible experimental searches for sphaleron production at RHIC/LHC
\cite{Shuryak:2021iqu}. The sphalerons are 3-dimensional, static and purely magnetic configurations, that 
minimize the energy functional. To quantify them, we will focus on two main variables.
Their	 topological Chern-Simons number, and their squared field size
	 \begin{equation} 
	 N_{CS}\equiv { \epsilon^{\alpha\beta\gamma} \over 16\pi^2}\int d^3x  \left( A^a_\alpha \partial_\beta A^a_\gamma +{1\over 3}\epsilon^{abc}A^a_\alpha A^b_\beta A^c_\gamma \right), \,\,\,\,
	 \label{eqn_Ncs}
	 	\rho^2\equiv { \int d^3 x \vec x^2 \vec B^2
	 		\over \int d^3 x  \vec B^2 }
	 \end{equation}
If those are kept constant, by adding pertinent Lagrange multipliers to the action,  we can find the energy and 
Chern-Simons number in parametric form
	 \begin{eqnarray}
	 	U_{\rm min}(\kappa, \rho)=(1-\kappa^2)^2{3\pi^2\over g^2\rho}, \,\,\, 
	 	{N}_{CS}(\kappa)=\frac 14 {\rm sign}(\kappa)(1-|\kappa|)^2(2+|\kappa|) 	
	 \end{eqnarray}	
where $\kappa=0$ corresponds to top of the barrier, known as   the ``sphaleron". 	 
	 
	 	\begin{figure}[h!]
	\begin{center}
		\includegraphics[width=6cm]{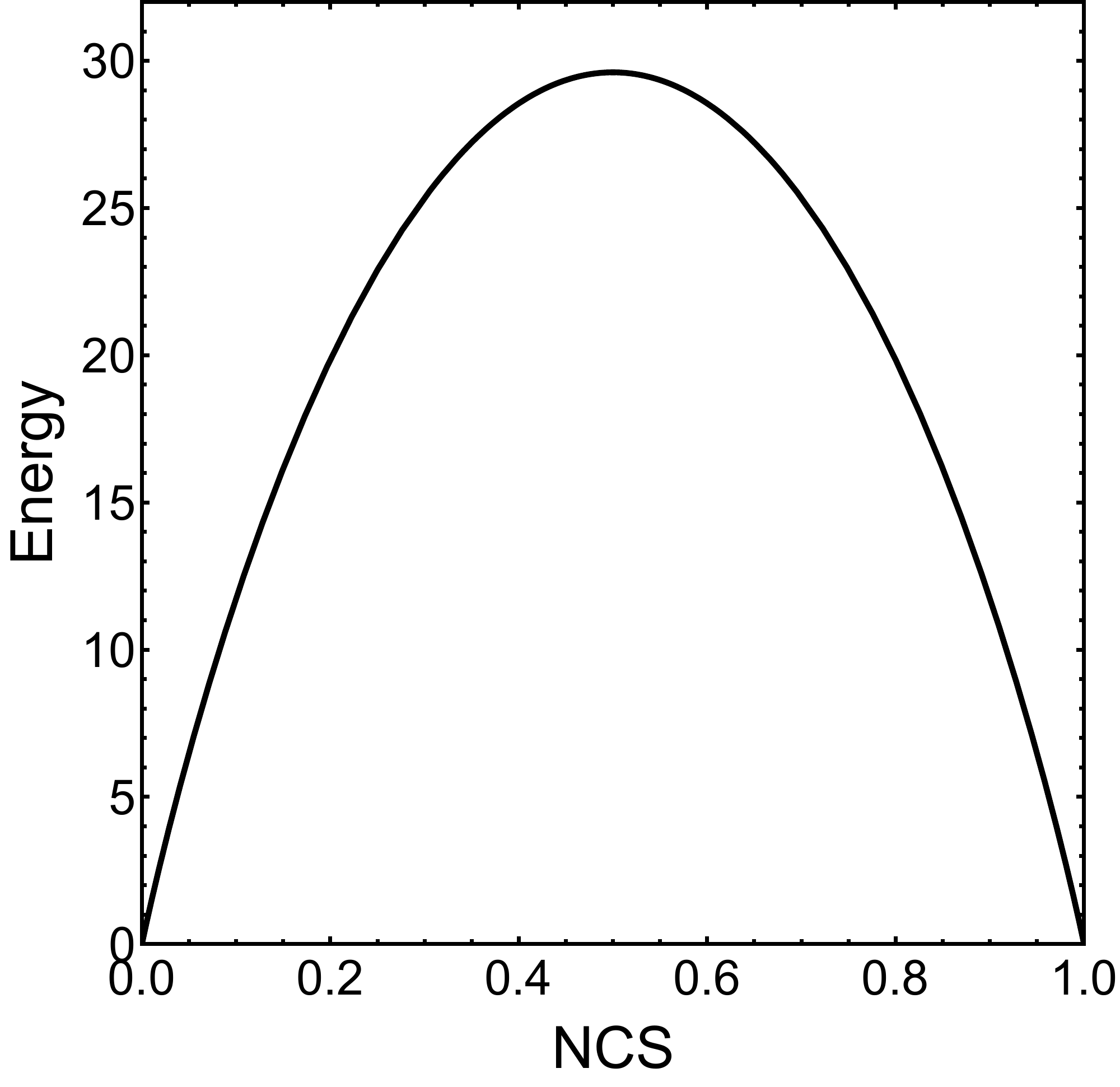}
		\caption{ 
			The potential energy $U_{\rm min}\bigg(N_{CS} ,\rho\bigg)$ (in units of $1/g^2\rho$)
			versus the Chern-Simons number ${N}_{CS}$  for the 
			``sphaleron path"  between $N_{CS}=0$ and $N_{CS}=1$.
		} \label{fig_sphaleron_path}
	\end{center}
\end{figure}

Production of sphaleron-path states can be described semi-classically using instanton-antiinstanton 
``streamline" 	\cite{Verbaarschot:1991sq,Khoze:1991sa,Shuryak:1991pn}, and their explosion in
Minkowski space-time by \cite{Shuryak:2002qz}. The new results consist in  estimates of how one
can produce QCD sphalerons,  as topologically-charged clusters in double-diffractive events, with
a cluster of few GeV mass at the center.  Its decay modes into 3 mesons were calculated using
$^\prime$t Hooft Lagrangian~\cite{Shuryak:2021iqu}.

\section{Mesons and baryon light-front wave functions and isospin asymmetry of the nucleon ``sea"} \label{sec-sea}
High energy processes, as in the famed electron-nucleon deep inelastic scattering, produced a rich
``parton phenomenology", in the form of parton distribution functions (PDFs),  distribution amplitudes (DAs), transverse momentum distribution  (TMDs) etc. 
They  are pertinent matrix elements of light front wave functions (LFWFs), which, are not well 
understood from first principles. 
In~\cite{Kock:2020frx,Kock:2021spt} light front DAs and PDFs were explored in the instanton liquid model using the
large momentum effective theory (LaMET) formulation~\cite{Ji:2013dva}, although the LaMET matching kernels may 
well be contaminated by non-perturbative small size instanton and anti-instanton contributions (also $\bar II$ molecules
as we discuss below)~\cite{Liu:2021evw}.

\begin{figure}[h]
\centering
\includegraphics[width=4cm,clip]{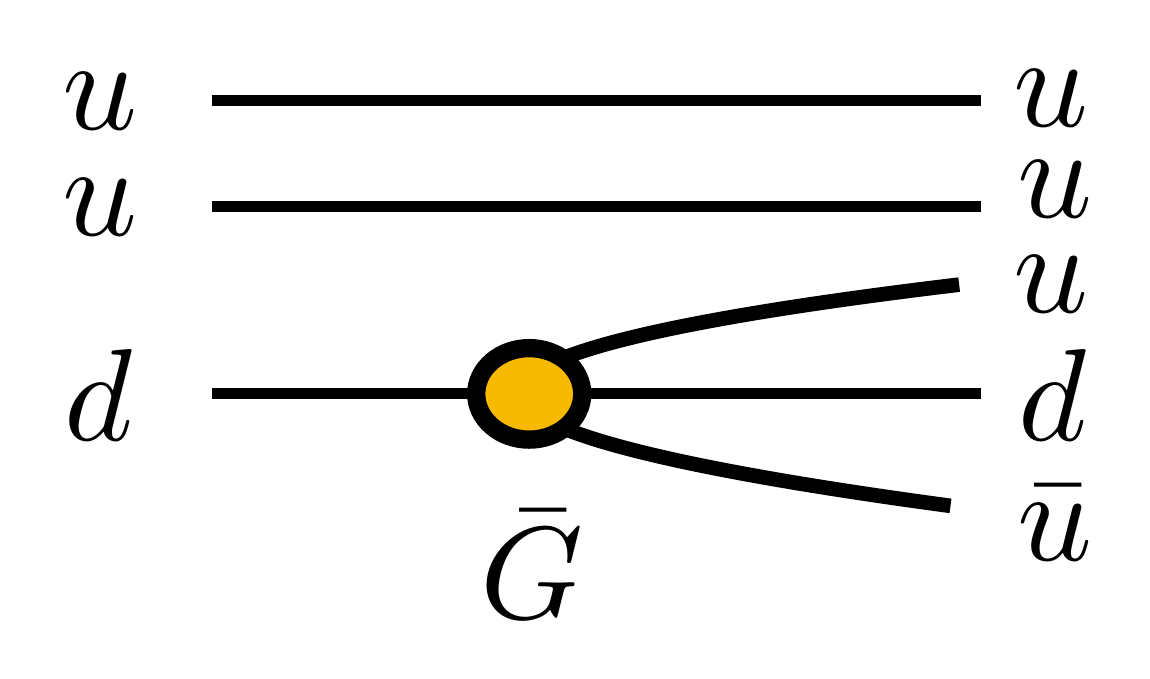}
\includegraphics[width=6cm,clip]{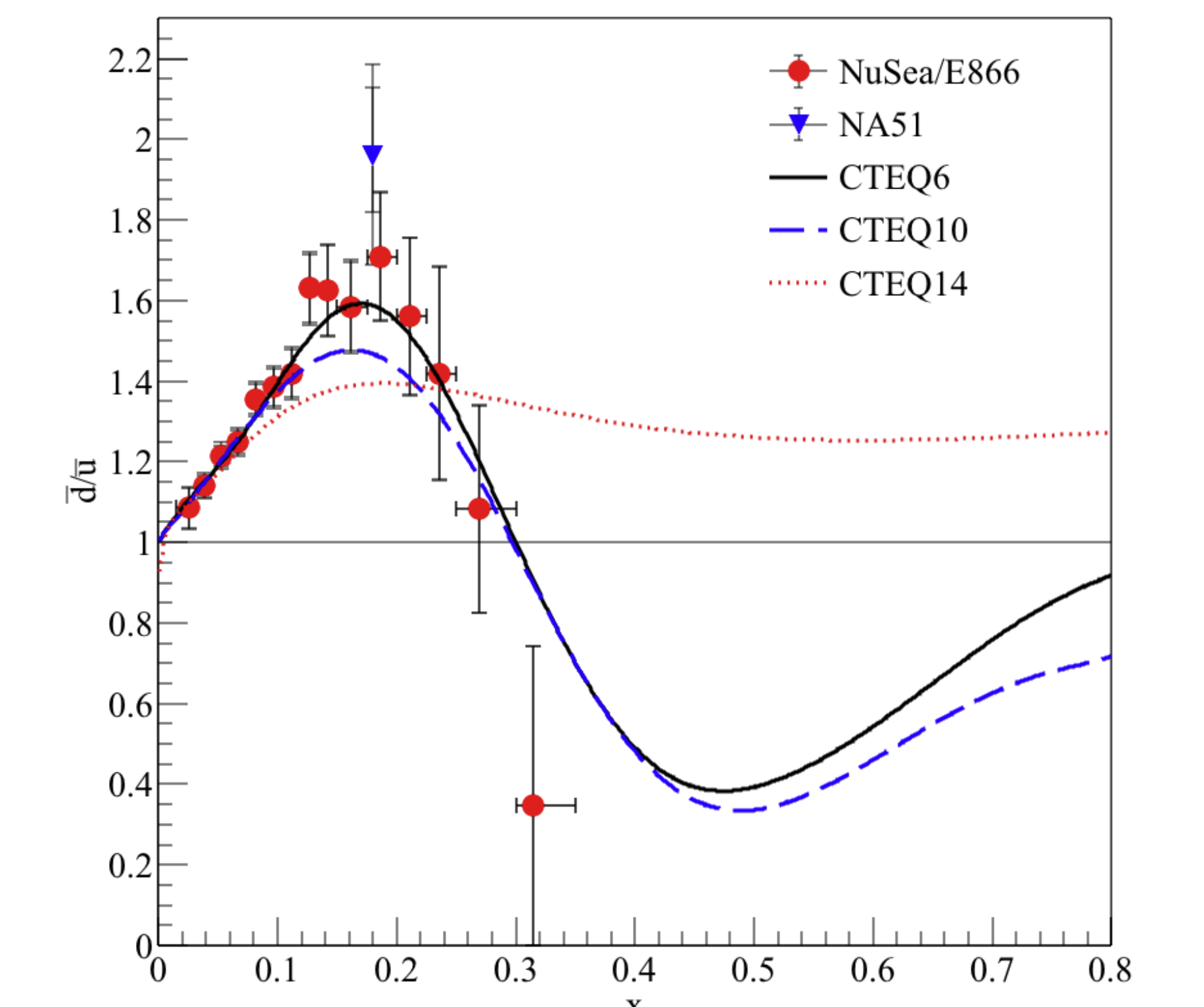}
\caption{Sea generation via 't Hooft Lagrangian (left) and the data on the $\bar d/\bar u$ ratio versus $x$ (right).} 
\label{fig-sea}       
\end{figure}

Furthermore,   the understanding of  the  light front Hamiltonians is only in its initial stage. 
In \cite{Shuryak:2019zhv}  LFWFs for several mesons and baryons were calculated, including  for the first time the 5-quark
component of the baryon, in relation  to the {\em isospin asymmetry puzzle}. The canonical process of quark pair production via gluons is ``flavor blind", and yet the antiquark sea of the nucleon is very asymmetric, as shown in Fig.\ref{fig-sea} (right).  As noted iin \cite{Dorokhov:1993fc}, the 't Hooft Lagrangian 
is  {\em flavor asymmetric}, say a $d$ quark can produce a $\bar u u$ pair (Fig.\ref{fig-sea} left) but $not$
a $\bar d d$. If it would be the only process, the $\bar d/\bar u$ ratio would be 2, as there are two
$u$ quarks and only one $d$ in the nucleon. This observation is not very far from the currently reported data for a given range of  $x$.

A calculation of the  admixture of the 5-q states to the nucleon LFWF, leads 
the right  magnitude for  the asymmetry and shape,  of  the first-generation antiquarks in agreement with the data.

\section{``Dense instanton ensemble"}\label{sec-dense}
The original instanton liquid model focused on the quark condensate and chiral symmetry breaking, and  therefore on the
stand-alone instantons, whose zero modes get collectivized. But the instanton ensemble also includes
 close instanton-antiinstanton pairs, or molecules. In so far, the application of the ``molecular component"   of the semiclassical ensemble was made
 only in connection to the phase transitions in hot/dense matter.
Indeed, this component is the only one which survives at
 temperatures $T>T_c$, where chiral symmetry is restored. Account for both components together
started with~\cite{Ilgenfritz:1988dh}. The ``molecular component"   was also shown to be
important at  high baryonic densities, where
it contributes to quark pairing and color superconductivity~\cite{Rapp:1997zu}.

\begin{figure}[h]
\begin{center}
\includegraphics[width=5cm]{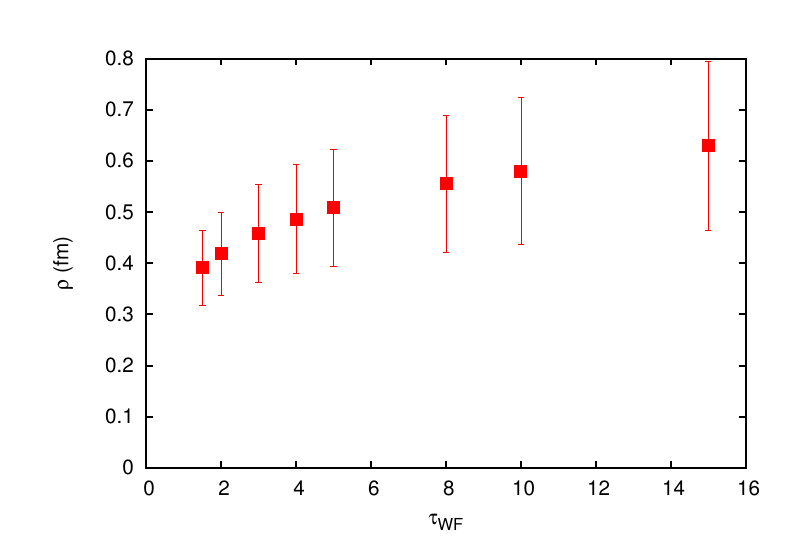}
\includegraphics[width=6cm]{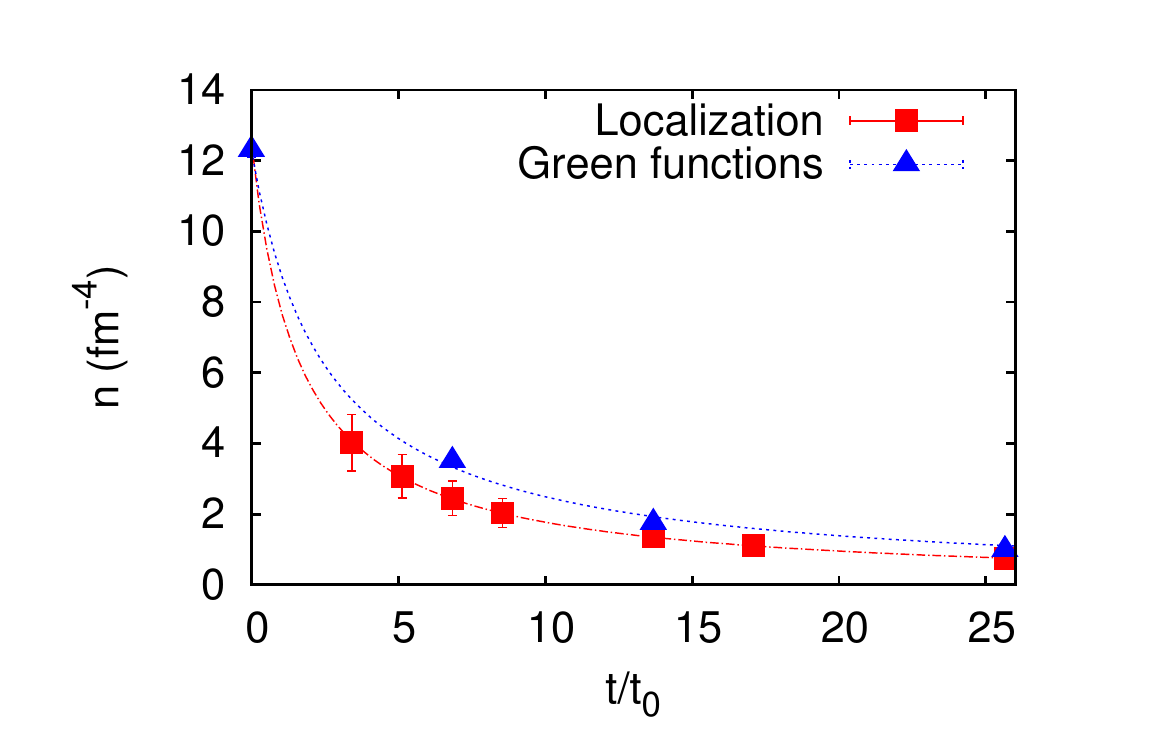}
\caption{The dependence of the mean instanton sizes (left  plot) and density (right  plot)
on the gradient flow cooling time $\tau=t/t_0$ (arbitrary units). The quantum vacuum 
 corresponds to an extrapolation to $\tau\rightarrow 0$.
}
\label{fig_cooling}
\end{center}
\end{figure}

 The  $\bar I I$  molecules were also observed on the lattice.  
The  stand-alone instantons are seen via ``deep cooling"  ,
during which the instanton-antiinstanton molecules get annihilated.
 The close $\bar I I$ pairs have been qualitatively studied in the recent 
work  in~\cite{Athenodorou:2018jwu},  which studied  their evolution during cooling, see Fig.\ref{fig_cooling}.
When extrapolated to zero cooling (left side of the plots),  one sees that while the instanton size fits previous
expectations (\ref{eqn_n_rho}), the density seems to actually be significantly larger.

\begin{figure}[h]
\centering
\includegraphics[width=6cm]{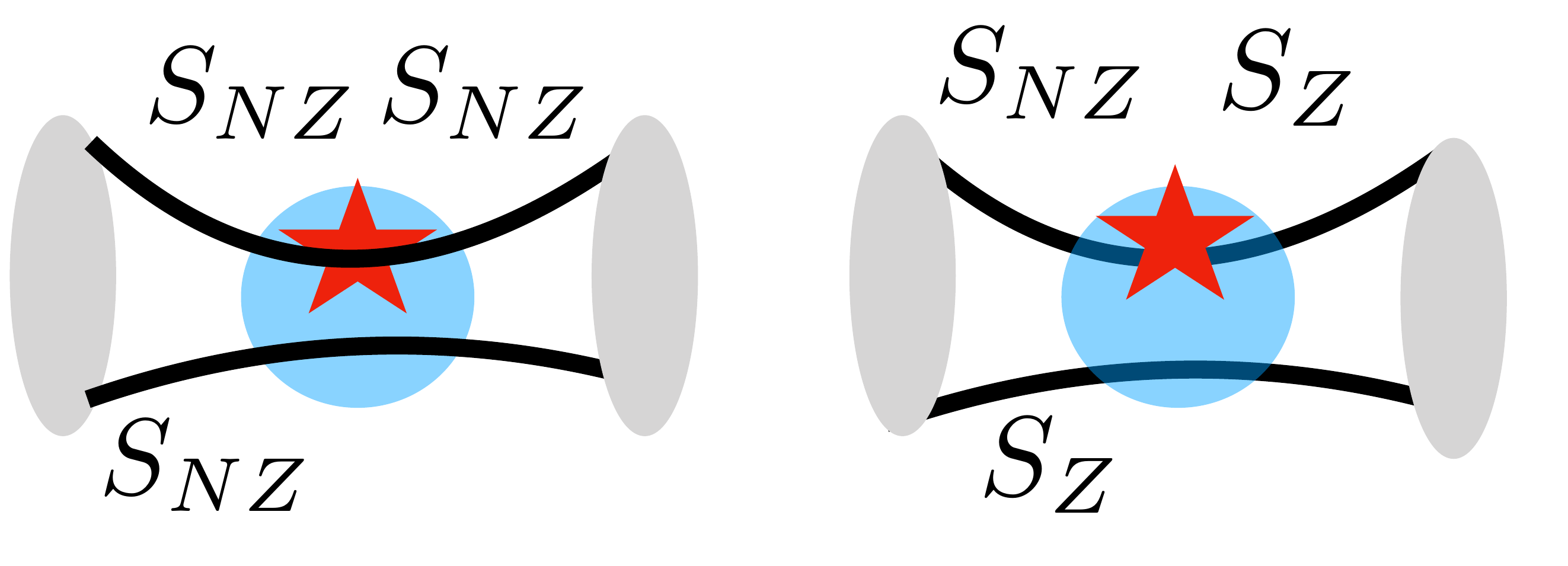}
\includegraphics[width=6cm]{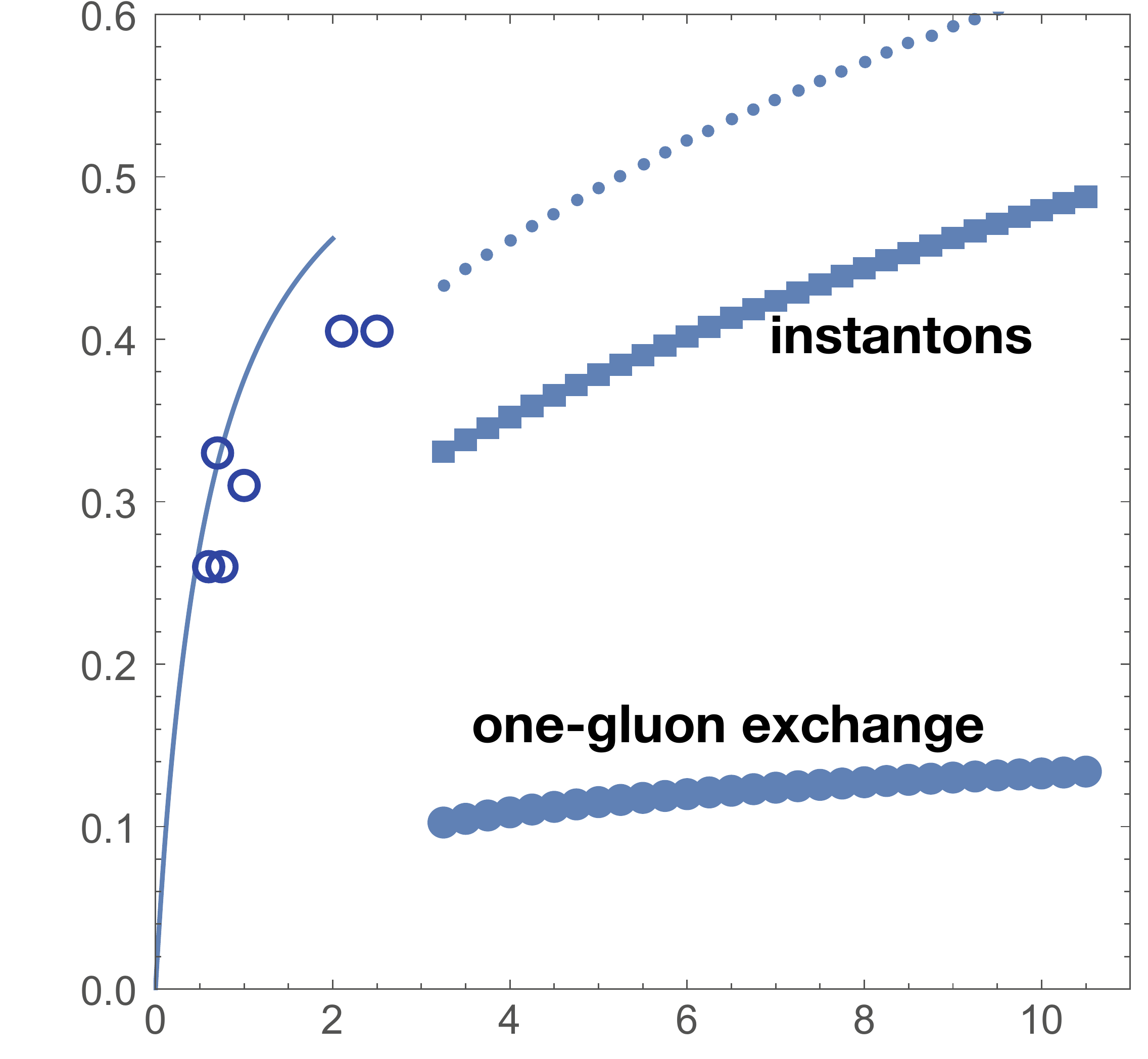}
\caption{Left: two diagrams with quark propagators in the instanton fields. Right: vector formfactor of the pion  $Q^2 F_\pi(Q^2)$. Open points correspond to some experimental data points, thin curve on the left
is standard dipole fit. The dotted line on the top is the sum of the perturbative and instanton contributions} 
\label{fig-ff}       
\end{figure}

\section{Formfactors: including instantons in the hard block} \label{sec-ff}
More recently,
we have explored  the non-perturbative contributions to the ``hard block" of mesonic form factors~\cite{Shuryak:2020ktq}.
We calculated the photon, scalar, graviton and dilaton FFs
 for the pion, rho and scalar $a_0$ (brother of $\eta^\prime$). The field at the instanton center is rather strong
$$ G_{rms}\equiv
\sqrt{ \big( G^a_{\mu\nu}(0) \big)^2 }=\sqrt{192}/\rho^2\approx 5 \, {\rm GeV}^2,$$ 
for a typical size of  $\rho=1/3\, fm$. This scale  
is comparable to $Q^2$ in the so called semi-hard region.
 
 One example, which was studied a lot experimentally, is the
 electromagnetic pion formfactor, shown in Fig.\ref{fig-ff} right.
It is dominated by  the diagram with three propagators made of non-zero Dirac modes (the left-most),
not the one with Dirac zero modes (in the middle). The contributions of the  ``dense instanton liquid" ,
through $\kappa=\pi^2 n_{I+\bar I} \rho^4\rightarrow 1$, is shown by the middle curve in  Fig.  \ref{fig-ff} right. 
The sum of the perturbative and instanton-induced formfactor reproduce the data. 
 
 The expressions
 for the perturbative, non-zero-mode and zero mode parts are
 \begin{eqnarray} \label{eqn_Vapi}
&&V^\pi_a(Q^2)=\epsilon_\mu(q)(p^\mu+p^{\prime\mu})\,(e_u+e_{\overline d})\,
\bigg[\bigg(\frac{2C_F\pi\alpha_sf_{\pi}^2}{N_cQ^2}\bigg) \nonumber\\
&&\times   \int  dx_1 dx_2 
\bigg({ 1\over \bar x_1\bar x_2+m_{\rm gluon}^2/Q^2}\bigg)\bigg( \varphi_{\pi}(x_1)\varphi_{\pi}(x_2)  \nonumber\\
&&+2{\chi_{\pi}^2 \over Q^2}    \bigg(
\varphi_{\pi}^P(x_1)\varphi^P_{\pi}(x_2)\bigg( {1 \over \bar x_2 +E_\perp^2/Q^2}-1\bigg)
+\frac 16\varphi_{\pi}^P(x_1)\varphi^{\prime \, T}_\pi(x_2)\bigg( {1 \over \bar x_2 +E_\perp^2/Q^2}+1\bigg)\bigg)\bigg)
\bigg]\nonumber\\
\end{eqnarray}
\begin{eqnarray} \label{eqn_Vcpi}
V_c^\pi= && \epsilon_\mu(q)(p^\mu+p^{\prime\mu})\,(e_u+e_{\overline d})\, \bigg[
{\kappa \pi^2f_{\pi}^2 \chi_\pi^2 \over N_cM_Q^2}
\langle\rho^2  {\mathbb G}_V(Q \rho) \rangle  \nonumber  \\
&&\times\int dx_1 dx_2 \overline x_1  
 \bigg(\varphi_{\pi}^P(x_1)\varphi_{\pi}^P(x_2)
-\frac 1{36}\varphi_{\pi}^{'T}(x_1)\varphi_{\pi}^{'T}(x_2)\bigg)\bigg]
\label{eqn_Vcpi}
\end{eqnarray}
\begin{eqnarray} \label{eqn_Vdpi}
&&V_d^\pi= -\epsilon_\mu(q)(p^\mu+p^{\prime\mu})\,(e_u+e_{\overline d})\, \nonumber\\
&&\times \bigg[\bigg(\frac {1}{N_c^2(N_c+1)}\bigg) \frac {4\kappa \pi^2f_\pi^2\chi_\pi^2}{3M_Q^2}\left<\rho^2\frac{K_1(Q\rho)}{Q\rho}\right>
\int dx_1 dx_2 \varphi_{\pi}^P(x_1)\varphi_{\pi}^{'T}(x_2) \bigg]\nonumber\\
\end{eqnarray}  
Here  $\varphi_{\pi}^P,\varphi_{\pi}^{T}$ are  the pion DAs
associated with the pseudoscalar $\gamma_5$ and tensor $\sigma^{\mu\nu}$ channels.
The leading twist $\varphi_{\pi}$ is the standard matrix element of the axial current $\gamma_\mu\gamma_5$. 
In the figure, all three DAs are taken to be just constant, independent of $x$. (Therefore
all  tensor DAs with an x-derivative vanish.)

\section{Instanton-induced inter-quark forces} \label{sec-forces}
Historically, hadronic spectroscopy got  a  solid foundation in 1970's, with the discovery of 
nonrelativistic quarkonia made of heavy $c,b$ quarks. In the first approximation, those are
well described by the simple Cornell potential
\begin{equation}  V_{\rm Cornell}(r)=-{4 \alpha_s \over 3}{1 \over r}+\sigma_T r 
\end{equation}
which correctly attributes the short-distance
potential to the perturbative gluon exchange, and its large distance ${\cal O}(r)$ contribution  to the tension of a confining flux tube (the QCD string). The issues to be discussed deal with  the nonperturbative origins of the inter-quark interactions  at $intermediate$ distances $r\sim 0.2-0.5 \,$fm.

\begin{figure}[h]
\centering
\includegraphics[width=6cm]{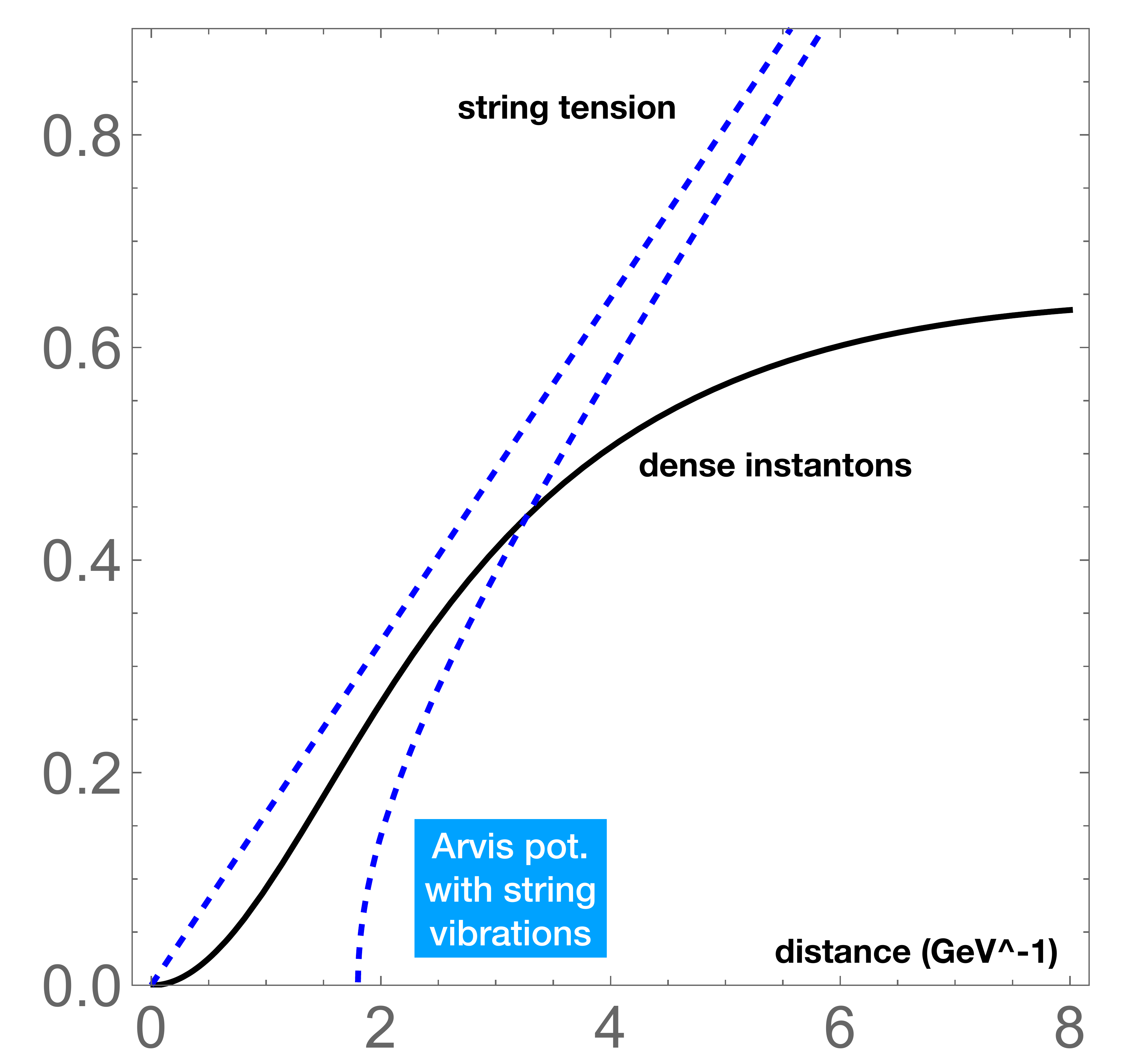}
\includegraphics[width=6cm]{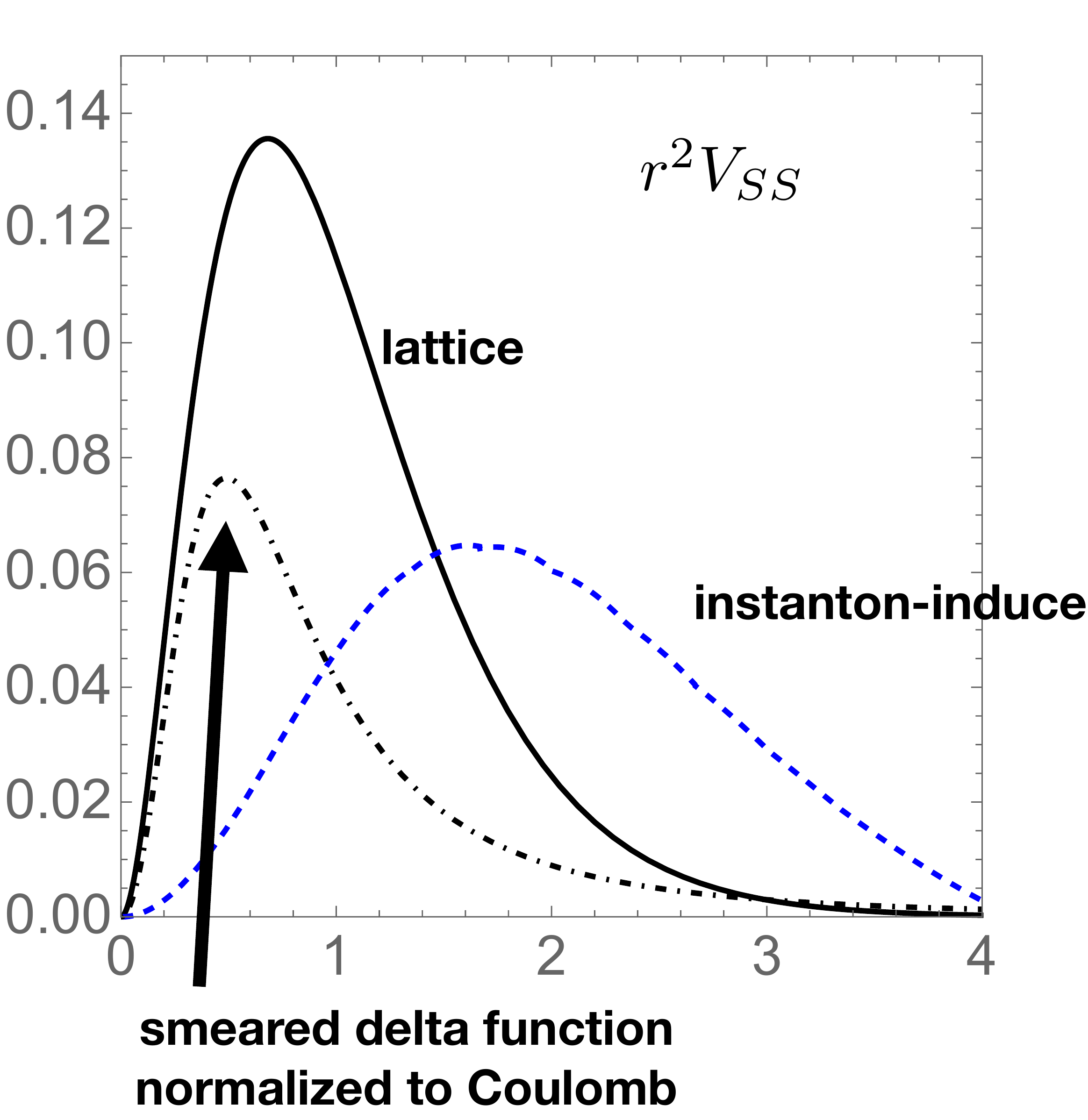}\\
\includegraphics[width=12cm]{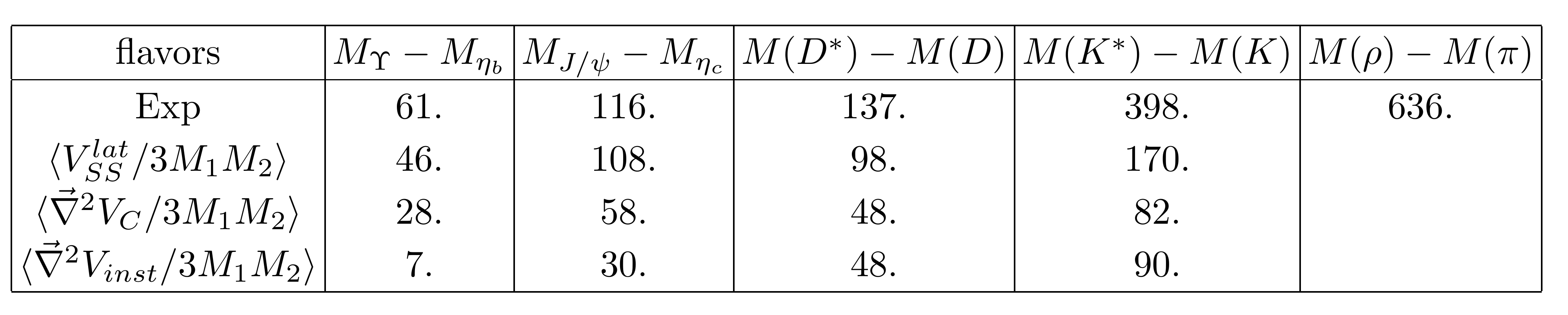}
\caption{The central potential  $V_c$ (left) and spin-spin $r^2V_{SS}(r)$ (right) versus distance $r, (GeV^{-1})$. 
The lattice result is one of the parameterization given in \cite{1508.02178}, the
perturbative is Laplacian of (regulated) Coulomb term.  } 
\label{fig-V}       
\end{figure}

Later developments  in~\cite{Callan:1978ye,Eichten:1980mw} connected  the static interquark potential,
 to the correlator of static Wilson lines (central)\begin{equation} e^{-V_C(r) T}=\langle W(\vec x_1) W^\dagger(\vec x_2)\rangle 
\end{equation} 
 The spin-dependent
forces were related to such  correlators with two magnetic fields ($V_{SS},V_{tensor}$), 
or  a magnetic  and electric field for the spin-orbit one. To evaluate such nonlocal quantities, one needs to use lattice simulations, or
rely on certain model of the vacuum fields. In Fig.\ref{fig-V} (left) we show the predictions from a  ``dense instanton model" to the central potential, compared to the  linear potential, and its version from~\cite{Arvis:1983fp},  including the string quantum vibrations (resummed ``Lusher terms").  One can see that instantons can complement smoothly, the flux tube
at intermediate distances.

One problem with the electric flux tube model is that it does not provide magnetic fields, while instantons are
self-dual and have $B$ (as large as $E$) badly needed to generate the spin forces. We calculated the instanton contributions to spin forces, see
$V_{SS}$, in  comparison to the  lattice potential, perturbative and instanton-induced in 
Fig.\ref{fig-V} (right). Note that the area below the perturbative and instanton-induced terms is comparable. 
The corresponding matrix elements (using the Cornell wave functions with proper quark masses)
is shown in the table. One can see that e.g. for charmonium,  the magntitude of the spin-spin term is
in agreement with the  lattice estimate,  and the level splitting. We also considered the $L=1$ families of mesons, 
from heavy to light, and considered other spin-dependent potentials. We also discussed $\bar I I$ molecules, 
which provide somewhat different potentials due to their different field content.

However for heavy-light and light-light cases this $V_{SS}$ is not enough.
The missing part is  attributed to the part of the quark propagators containing zero modes ($^\prime$t Hooft Lagrangian). It works well for heavy-light
and the pions, as expected.

\section{Hadrons on the light front}

\begin{figure}
		\includegraphics[width=6cm]{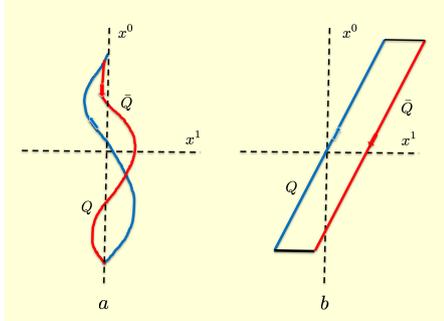}
		\caption{$\bar Q Q$ meson in the rest frame (a) and in the light-front frame (b).}
		\label{fig_qqbar}
\end{figure}

In the rest frame, the non-relativistic description of hadrons and their constituents is most
natural for heavy quarks, with the use of non-relativistic kinectic energy and potentials. 
However, this  approach is not justified for light quarks, as masses and transverse momenta
are comparable. Their
quantum motion is complicated as illustrated in Fig.~\ref{fig_qqbar} (left). The situation however 
is much more ``democratic" on the light front, where all constituents motion are frozen anyway
in Fig.~\ref{fig_qqbar} (right). All interactions can be deduced from the pertinent Wilson
loops, modulo tunneling due to zero modes.

\subsection{Light front Hamiltonian from the QCD vacuum}

Indeed, let $Q\bar Q \equiv Q_1Q_2$ be the world-lines assigned to the  Wilson lines
composing a meson in Fig.~\ref{fig_qqbar} (right). In Euclidean signature, the Wilson lines
are sloped at an angle $\theta$, and their light-like limit follows by analytical continuation 
$\theta\rightarrow -i\chi$  \cite{Meggiolaro:1997mw,Shuryak:2000df,Giordano:2009su,Shuryak:2021hng}.
For the squared meson mass operator, or light front Hamiltonian $H_{LF}$, the result~\cite{Shuryak:2021hng}

\bea
\label{16Z} 
H_{LF}\approx && \frac{k_\perp^2+m_Q^2}{{x\bar x}} + 2P^+P^-\nonumber\\
\approx &&
 \frac{k_\perp^2+m_Q^2}{{x\bar x}} 
+ 2M (\mathbb V_{Cg}(\xi_x)+\mathbb V_C(\xi_x)
+\mathbb V_{SD}(\xi_x, b_\perp)
 +\mathbb V_{TH}(\xi_x, b_\perp)) \nonumber\\
\eea
On the light front, the  invariant distance $\xi_x$ is

\bea
M\xi_x=\big(\big|{id/dx}\big|^2+M^2{b_\perp}^2\big)^{\frac 12}
\eea
The  longitudinal distance $\gamma b_3=id/dx/M$, is the  conjugate of Bjorken-x or $x=k^3/P^3$.

The one-gluon exchange potential $V_{Cg}$ in the instanton vacuum is
\bea
\label{ONEG}
\mathbb V_{Cg}(\xi_x)=-\frac{g^2 T_1^AT_2^A}{2\pi^2}  \frac 1{\xi_x}\int_0^\infty\frac {dx\,x{\rm sin}x}{x^2+(\xi_xm_G(x\rho/\xi_x))^2}         
\rightarrow -\frac{g^2 T_1^AT_2^A}{4\pi}\frac {e^{-m_G \xi_x}}{\xi_x}\nonumber\\
\eea
with the running gluon mass~\cite{Musakhanov:2021gof}

\bea
m_G(k\rho)&=&m_G\, \bigg(k\rho\,K_1(k\rho)\bigg)\nonumber\\
 m_G\rho &\approx& 2\bigg(\frac {6\kappa }{N_c^2-1}\bigg)^{\frac 12}\approx 0.55
\eea
using the estimate $\kappa=\pi^2\rho^4 n_{I+\bar I}$ in the right-most result.

The instanton-induced central potential $V_C$  is

\be
\label{CENT1}
\mathbb V_C(\xi_x)= \bigg(\frac{4\kappa }{N_c \rho}\bigg){\bf H}(\tilde\xi_x)
\ee
with  the integral operator 
\bea
\label{W10X}
{\bf H}(\xi_{x} )=&&\int_0^\infty y^2 dy\int_{-1}^{+1} dt \nonumber\\
&&\times\bigg[1-{\rm cos}\bigg(\frac{\pi y}{\sqrt{y^2+1}}\bigg)
{\rm cos}\bigg(\pi\bigg(\frac{y^2+\tilde\xi_x^2+2\xi_{x}yt}{y^2+\tilde\xi_x^2+2\tilde\xi_{x}yt+1}\bigg)^{\frac 12}\bigg)\nonumber\\
&&-\frac{y+\xi_{x}t}
{(y^2+\xi_x^2+2\tilde\xi_{x}yt)^{\frac 12}}
{\rm sin}\bigg(\frac{\pi y}{\sqrt{y^2+1}}\bigg)
{\rm sin}\bigg(\pi\bigg(\frac{y^2+\tilde\xi_x^2+2\tilde\xi_{x}yt}{y^2+\xi_x^2+2\tilde\xi_{x}yt+1}\bigg)^{\frac 12}\bigg)\bigg]\nonumber\\
\eea
with the dimensionless invariant distance on the light front $\tilde\xi_x=\xi_x/\rho$. ${\bf H}(\tilde\xi_x)$ admits the short  distance limit

\bea
\label{17Z} 
&&{\bf H}(\tilde\xi_x) \approx + \bigg(\frac{\pi^3}{48}-\frac{\pi^3}3J_1(2\pi )\bigg)\tilde\xi_x^2\nonumber\\
&&+
\bigg(-\frac{\pi^3(438+7\pi^2)}{30720}+\frac{J_2(2\pi)}{80}\bigg)\tilde\xi_x^4
\eea
and large distance limit
\bea
{\bf H}(\xi_x) \approx -\frac{2\pi^2}3\bigg(\pi J_0(\pi)+J_1(\pi)\bigg)+\frac{C}{\tilde\xi^p_x}
\eea
with $p\ll 1$ and $C>0$. The  large asymptotic is to be subtracted in the definition of the potential. 
In the dense instanton vacuum~\cite{Shuryak:2021hng}, the central potential (\ref{CENT1}) is almost linear 
at intermediate distances $0.2-0.5$ fm, and is expected  to be taken over by the linearity of the confining 
potential at larger distances.

The spin potentials receive contribution from both the perturbative and non-perturbative parts of the underlying gluonic
fields. In particular, the non-perturbative instanton-induced spin potentials can be derived in closed form
\bea \label{eqn_inst_spin_orb}
\mathbb V_{SD}(\xi_x, b_\perp)=&&2M\bigg(
\bigg(\frac{l_{1\perp}\cdot S_{2\perp}}{m_{Q1}m_{Q2}} -\frac{l_{2\perp}\cdot S_{1\perp}}{m_{Q1}m_{Q2}}\bigg)
\frac 1{\xi_x}\mathbb V_C^\prime(\xi_x)\nonumber\\
&&+\frac 1{m_{Q1}m_{Q2}}(S_{1 \perp}\cdot \hat{b}_\perp S_{2\perp} \hat{b}_\perp-\frac 12 S_{1\perp}\cdot S_{2\perp})
\frac {2b_\perp^2}{\xi_x}\mathbb V_C^{\prime\prime}(\xi_x)\bigg)\nonumber\\
\eea
with the respective spins $\vec S_{1,2}=\vec \sigma_{1,2}/2$, and transverse  orbital momenta

\bea
\label{PERP2}
l_{1,2\perp}=\pm (b_\perp\times m_{Q1,2}s_{1,2}\hat 3)_\perp\qquad 
 s_{1,2}={\rm sgn}(v_{1,2}) \rightarrow \frac{Mx_{1,2}}{m_{Q1,2}}
 \eea
The contributions stemming from the zero modes $\mathbb V_{TH}(\xi_x, b_\perp)$ are not included. They follow
from the $^\prime$t Hooft determinantal interaction properly continued to the light front.

\subsection{Light front spectra and wavefunctions}

We have used  the light front Hamiltonian (\ref{16Z}) to analyze heavy and light mesons and baryons.
One strategy consists at linearizing the confining part of the potential (instanton induced at small distances, and
string induced at large distances), and treating the Coulomb and spin contributions perturbatively. More specifically, 
the central part of the potential can be linearized using the ``einbein trick" 
\bea
\label{EINBEIN}
2M \mathbb V_C(x,b_\perp)=2\sigma_T\big(|id/dx|^2+M^2 b_\perp^2\big)^{\frac 12}\rightarrow \sigma_T\bigg(\frac{|id/dx|^2+b\, b_\perp^2}a+a\bigg)
\eea
with  $a,b$  variational parameters.
The minimization with respect to $a$ would lead back to a previous expression, but the trick is to
do it $after$ the Hamiltonian is diagonalized.
It is also followed by the substitution 
$b\rightarrow M^2\approx (2m_Q)^2$ for heavy mesons,  and most light ones. 
The light front Hamiltonian is then 
\be 
\label{HLFX}
H_{LF}=H_0+\tilde V + V_{perp}+V_{spin} 
\ee
with
\begin{equation}
\label{H0X}
 H_0={\sigma_T \over a} \bigg( -{\partial^2  \over \partial x^2}-b{\partial^2  \over \partial  \vec k_\perp^2} \bigg) + \sigma_T  a + 4(m_Q^2+ k_\perp^2)
\end{equation} 
and $\tilde V$
\begin{equation}  
\tilde V(x,\vec k_\perp)\equiv (m_Q^2+k_\perp^2)\bigg({1 \over x \bar x} -4\bigg)
\end{equation}
where $\bar x=1-x$. $H_0$ is diagonal in the functional basis used~\cite{Shuryak:2021hng}. 
We use the momentum representation, with $\vec k_\perp$ as variable. 
The residual interaction has non-zero matrix elements   $\langle n_1 |V(x,\vec k_\perp) | n_2\rangle $ for all $n_1,n_2$ pairs. 
The perturbative part $V_{perp}$ for heavy quarks is the Coulomb term, with 
running coupling and other radiative corrections. Finally, the last term $ V_{spin} $
contains matrices in spin variables and in orbital momenta.

\begin{figure}[t]
\begin{center}
\includegraphics[width=6cm]{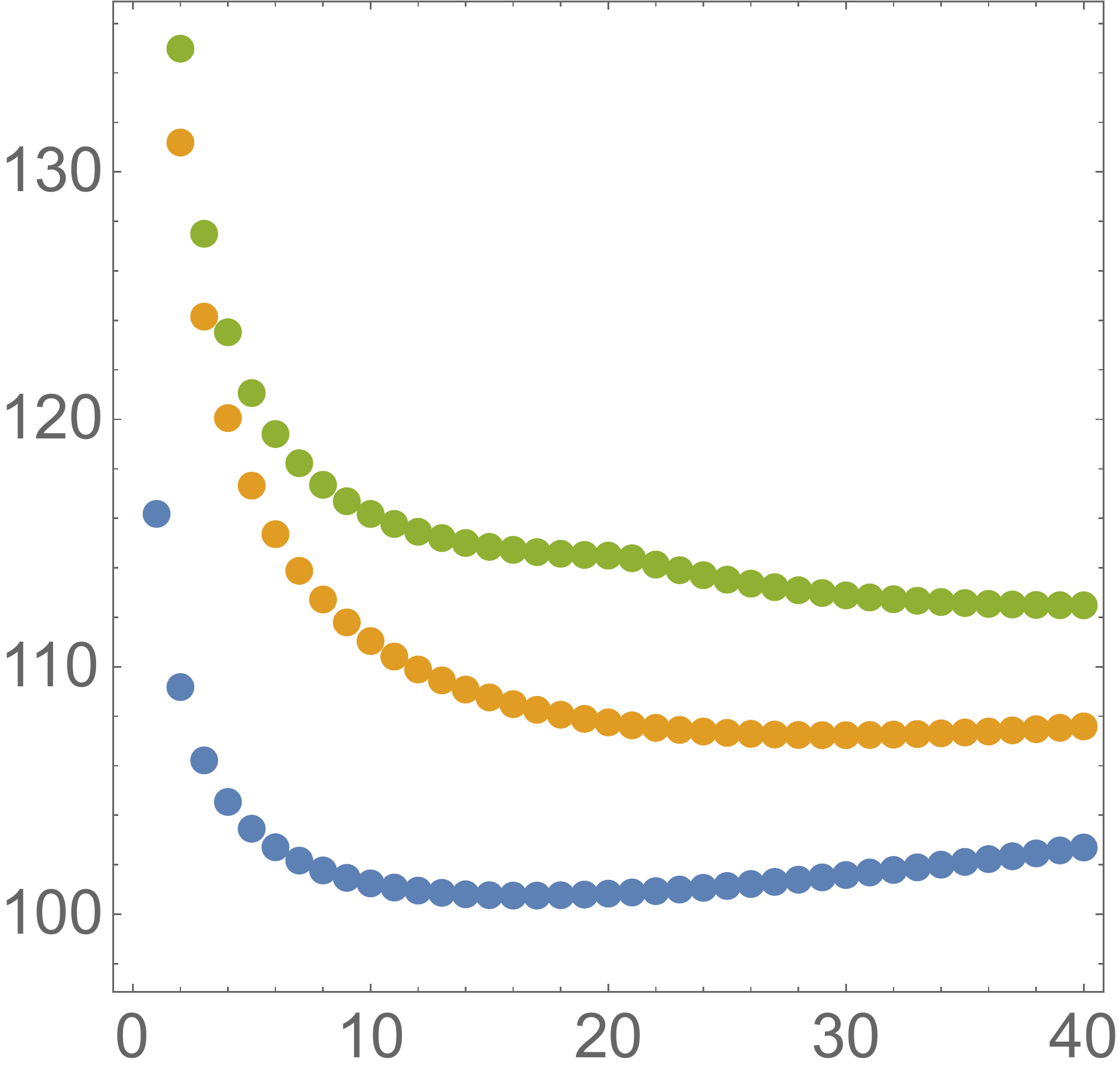}
\includegraphics[width=6cm]{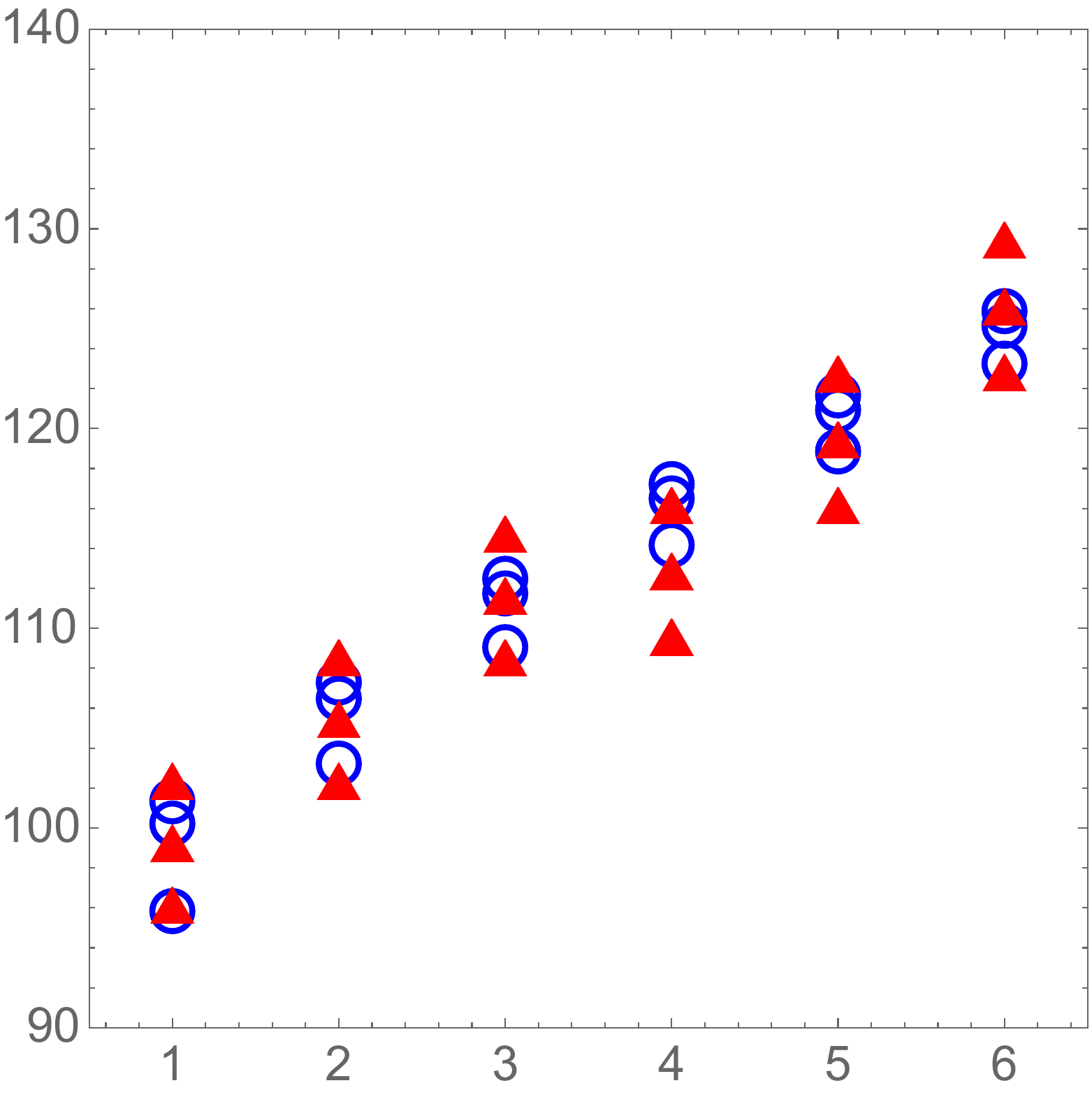}
\caption{Top: Squared masses $M^2_{n}$ for  $\bar b b$ mesons for $n=1,2,3$ versus the variational parameter $a$.
Bottom:  Squared masses  for $n=0..5$ (left to right) and orbital momentum 
$m=0,1,2$ (down to up), calculated from the light front Hamiltonian $H_{LF}$ (red triangles),
and shifted by a constant, $M^2_{n+1}-5\, GeV^2$. For comparison, the blue-circles show the squared masses $M^2_{n+1}$ calculated from Schroedinger equation in the CM frame,  with only  linear plus centrifugal potentials.}
\label{fig_M2_vs_a}
\end{center}
\end{figure}

 To solve the problem we have two numerical strategies. First we used a (truncated) basis set and write the Hamiltonian
in the form of a 12$\times$ 12 matrix, and diagonalize $H_0+\tilde V$, to find its eigenvalues as a function of
the remaining parameter $a$. The results for the three lowest states 
$n=1,2,3$ are shown in Fig. \ref{fig_M2_vs_a} (left).  The near flateness in $a$ around the minima,
suggest $a=25$ as a common value. This procedure preserves the orthonormality of the eigen-spectrum.

Another approach worked out later is direct numerical solution in 3-d space of transverse momenta and $x$.
To our delight, the wave functions found in both methods agree within the width of the lines.

\begin{figure}[htbp]
\begin{center}
\includegraphics[width=6cm]{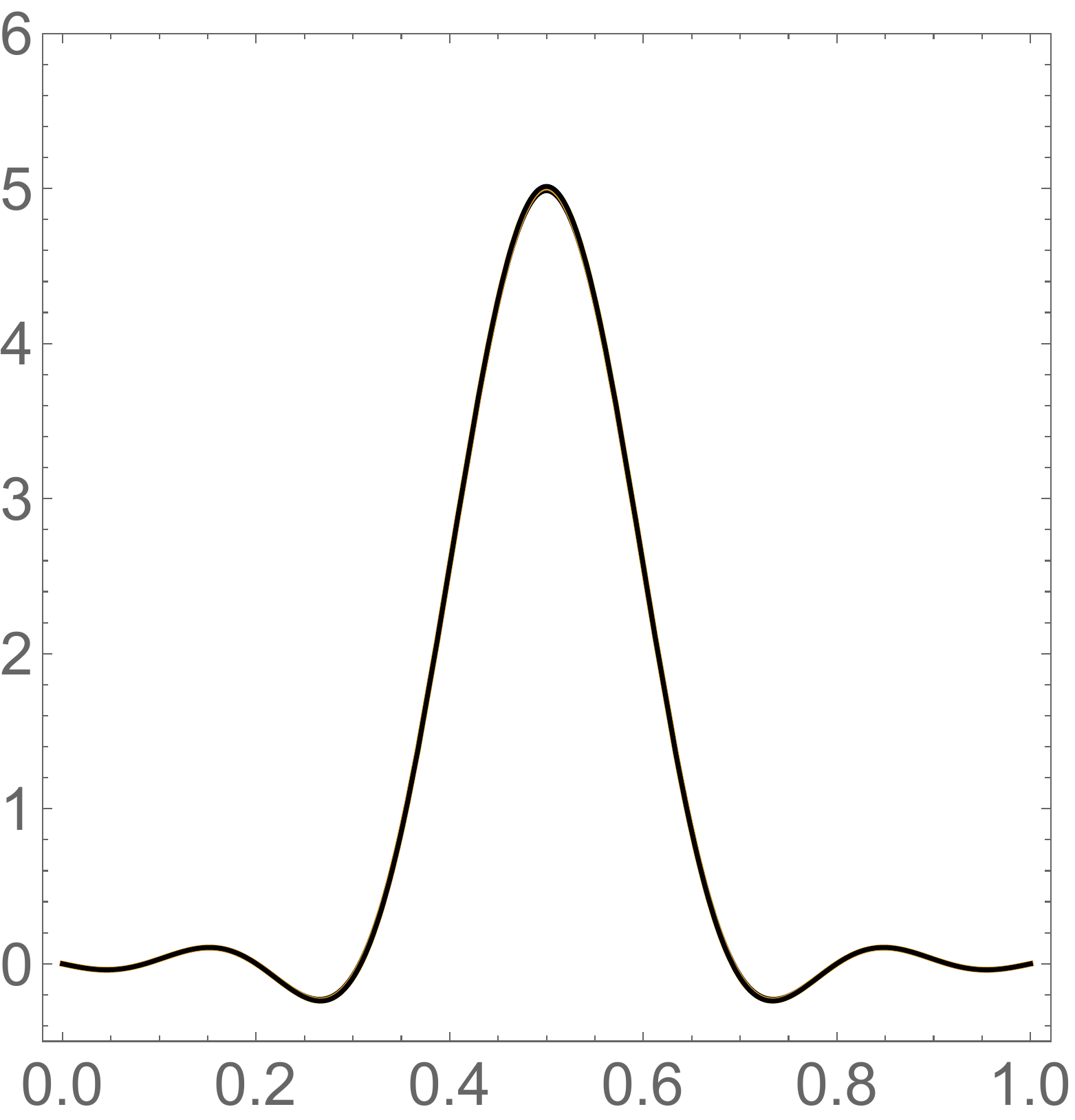}
\includegraphics[width=6cm]{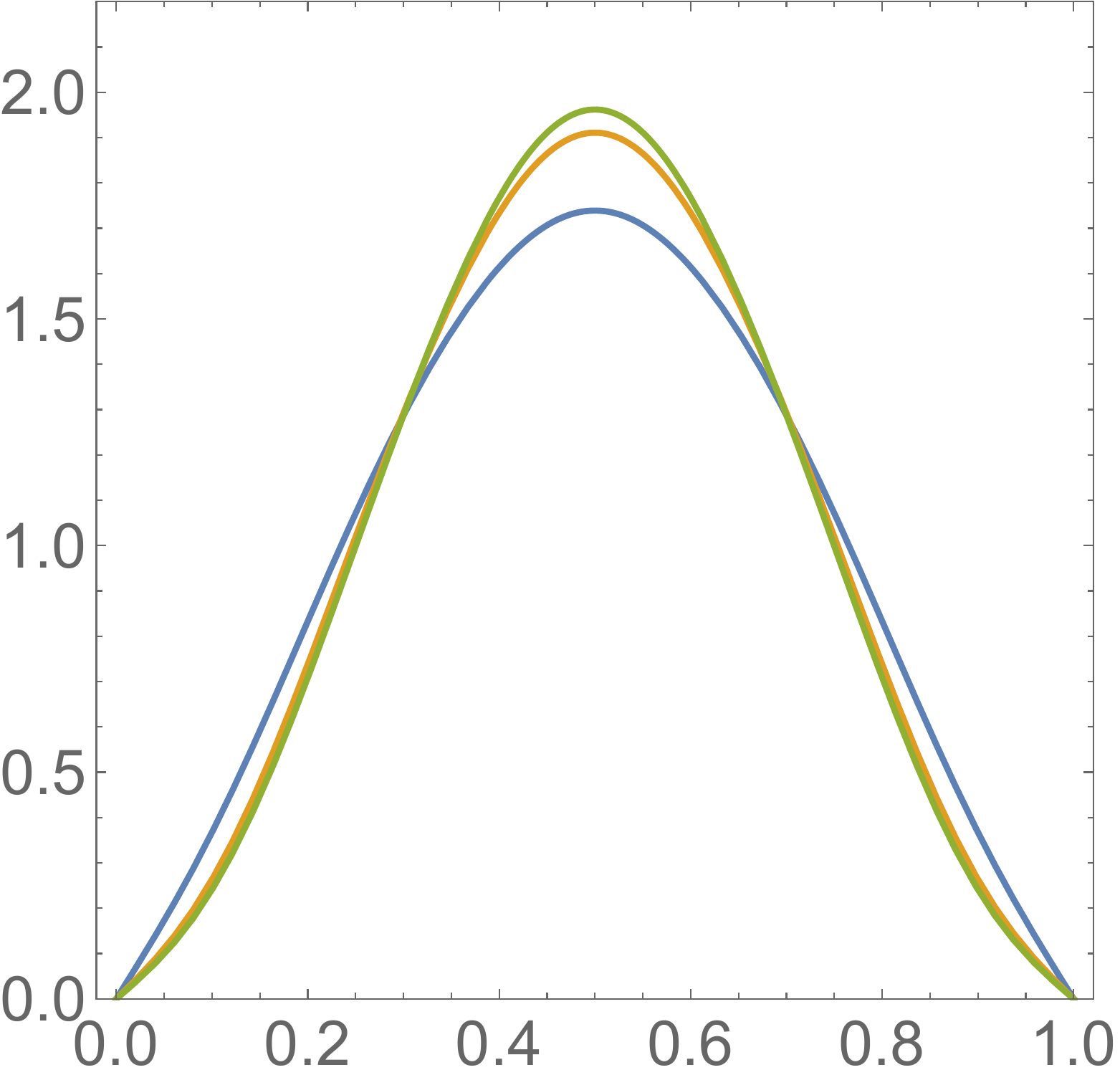}
\caption{ Distribution amplitudes for $\bar b b$ (upper) and a  ``generic" light $\bar q q$ meson (lower) as a function of $x$,
for the three lowest states $n=0,1,2$. For bottomium, the difference between the  three curves  is too small to be
visible. For the  light meson, the differences are visible.
With  increasing $n$, the DAs become narrower and higher at
$x=\frac 12$.}
\label{fig_3DAs}
\end{center}
\end{figure}

The  calculated masses (shifted by a constant  ``mass renormalization", to make $n=0,m=0$ states the same) are shown in  Fig.\ref{fig_M2_vs_a} (right).
The bottom part shows  good agreement between the masses  obtained solving  the Schroedinger equation 
in the rest frame (blue circles),  and the masses following from the  light-front frame (red-triangles).
The slope is correct, and is determined by the same string tension $\sigma_T$. The splittings in orbital momentum are of the same scale, but not identical.
This is expected, as we compare the 2-dimensional  $m$-states on the light front,  with the 3-dimensional  $L$-states in the center of mass frame.
The  irregularity between the third and  fourth set of states, is due to our use of a 
 modest  basis set, with only three radial functions  (altogether 12 functions if one counts them with 4 longitudinal harmonics). 
 This can be eliminated using a larger set. The corresponding wavefunction for bottomium is shown in Fig.~\ref{fig_3DAs} (left), 
 and for a typivcal light meson $\bar q q$ is shown Fig.~\ref{fig_3DAs} (right). Further details can be found in
 our subsequent papers.

\section{Semiclassical theory of deconfinement and chiral phase transitions based on instanton-dyons}
Instanton-dyons are topological solutions of YM equations at finite temperatures.
Their semiclassical ensembles were studied by a number of methods, including 
direct Monte-Carlo simulation, for SU(2) and SU(3) theories, with and without fermions.
We present these results and compare them with those from lattice studies. We also
consider two types of QCD deformations. One consists in adding operators with powers of 
the Polyakov line, affecting deconfinement. Another consists in changing the quark periodicity condition, affecting the chiral transition. 
We will also discuss how lattice configurations (with realistic quark masses) can be used 
to unravel the zero and near-zero Dirac modes of the underlying gauge configurations. 
The results are in  good  agreement with the analytic instanton-dyon theory. In sum, 
the  QCD phase transitions are well described in terms of such semiclassical configurations.

At finite temperature the {\em Polyakov line} has certain nonzero expectation values,
or $holonomies$ $\mu_i(T)$ (see below). The instanton solution deformed by an  asymptotics $A_0$ field,  splits into instanton constituents, known as {\em instanton-dyons} or {\em instanton-monopoles}
Like instantons, they have topological charges. Unlike instantons, those are
{\em not quantized to integer} $Q$, which is possible because they are wired by Dirac
strings due to their magnetic charges.

Remarkably, the partition function for monopoles follows from that of instanton-dyons
by {\em Poisson duality}, which is tightly related to Hamilton and Jacobi duality when
describing dynamical systems.
More details regarding this, and further discussions on the
QCD flux tubes are in~\cite{Shuryak:2021vnj}.


\subsection{Instanton-dyons on the lattice}

A thorough discussion of the  Kraan-van Baal solution for an instanton,
containing $N_c$ instanton-dyons, or  the formalism leading to their zero modes,
is not possible in this short review, and we refer to the original papers for that.
For completeness, we recall that the eigenvalues of Polyakov line, designated as $\mu_i(T), i=1..N_c$ can
be regarded  as locations on a unit circle. Their differences $\nu_i=\mu_{I+1}-\mu_i$ 
are lengths of the corresponding fractions of the circle, 
$S_i=\nu_i S$
where
$$S=8\pi^2/g^2=\bigg({11\over 3}N_c-{2\over 3} N_f\bigg) log\bigg({T \over \Lambda_{QCD}}\bigg) $$
Also, if quarks are given different periodicity phases $z_f, f=1..N_f$ over the Matsubara circle, the normalizable physical zero mode belongs to the dyon
for which  $z_f$ belongs to the sector  $z_f\in [\mu_{i},\mu_{i+1}]$.  So, using
different $z_f$ one can see all dyon types. 

Let us start with the description of the  efforts to
identify these objects on the lattice. The key method is ``cooling" 
 of vacuum configurations, which was used efficiently to identify ensembles  of instantons in the 1990's.
The first lattice observation of instanton-dyons was via ``constrained minimization" by Langfeld and Ilgenfritz~\cite{Langfeld:2010nm}, 
fixing  $\langle P \rangle$, in which selfdual clusters with non-integer topological charges were seen. 
Gattringer et al and Ilgenfritz et al have since introduced and refined the ``fermionic filter"
allowing to identify the instanton-dyons via distinct zero modes and variable periodicity phases.

\begin{figure}[htbp]
\begin{center}
\includegraphics[width=7cm]{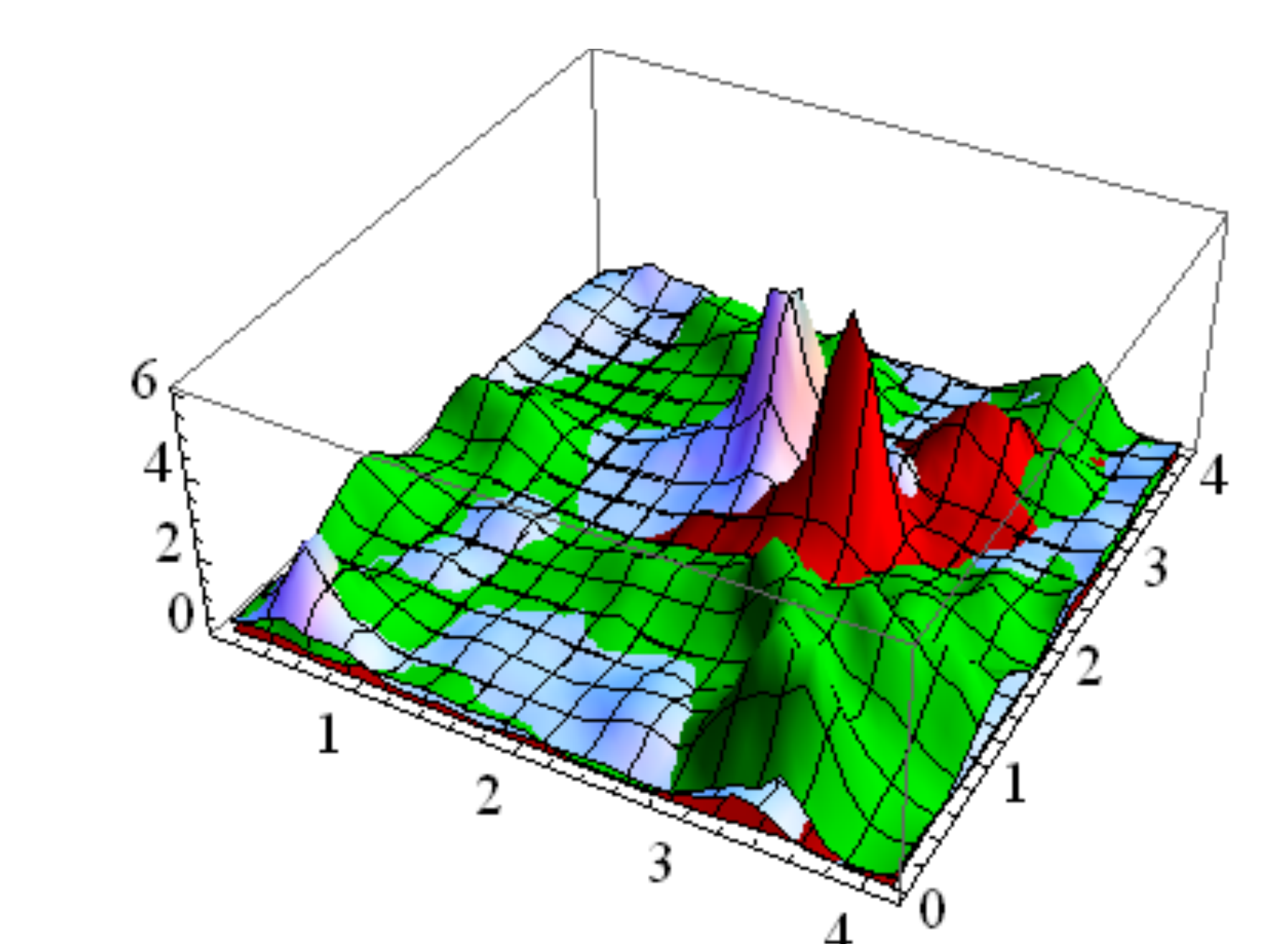}
\includegraphics[width=7cm]{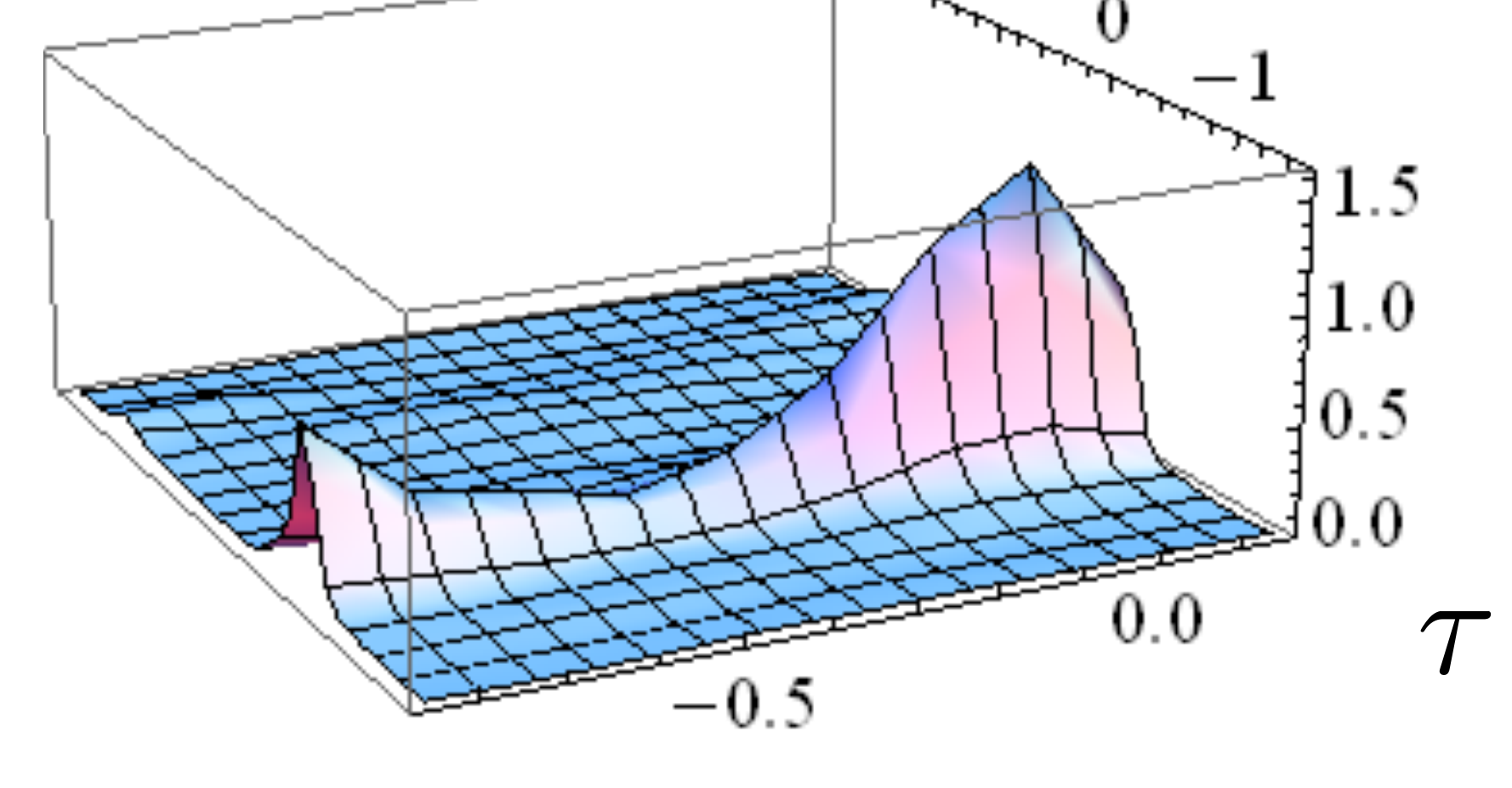}
\caption{(Top) Space slice of density of an exact zero mode from QCD simulaiton at $T=T_c$. The three colors
refer to dyons of three different types. (Bottom) Tau dependence of a dyon, perturbed 
by nearby dyons. See text.}
\label{fig_dyons_from_zero_modes}
\end{center}
\end{figure}

Some of the recent progress in these directions can be found 
in~\cite{Larsen:2018crg,Larsen:2019sdi}.  QCD simulations with realistic masses were
performed at and near $T_c$, using domain wall fermions with good chiral symmetry.
Using overlap fermions with exact chiral symmetry,  the focus was  on exact 
zero modes (and near-zero ones). The top part in~Fig.\ref{fig_dyons_from_zero_modes}
shows a typical landscape of the zero mode densities. There are three different dyon types  for  $N_c=3$.
The shape of isolated peaks are well described by analytic formulae from van Baal and collaborators, and 
derived for a single dyon. They are apparently undisturbed by 
millions of perturbative gluons present  in the ensemble.

Previous works however have not analyzed the ``topological clusters", the situations
in which two or three dyons overlap strongly. The  Kraan-van Baal solution
allows these cases, and  good agreement  was also found in the numerical
analysis of instanton-dyon ensembles in~\cite{Larsen:2018crg,Larsen:2019sdi}.
The right  part in~Fig.\ref{fig_dyons_from_zero_modes},
is an example of a (Euclidean time) $\tau$-dependence of 
the density. An isolated dyon should show no such dependence at all, and what
is seen is a result of an interference with overlapping dyons. Locating those and
using analytic expressions for the zero mode density,  confirms their identity.
The  {\em semiclassical description of zero and near-zero Dirac modes on the lattice is quite accurate, at least in terms of the 
zero mode shapes.}

\subsection{Studies of  instanton-dyon ensembles}
The simplest limiting case is weak coupling or very high temperature ($T\rightarrow \infty$), in which the
dyon density is exponentially small, with their interactions and back reactions  negligible. In QCD with $N_f$  quarks,  all twist factors are equal with $z_f=\pi$,
 and all the $L$ dyons combine 
in a  $^\prime$t Hooft  vertex with $2N_f$ legs. For $N_f=1$,  $U(1)_A$
chiral symmetry is explicitly broken at any  $T$, with exponentially small $\langle \bar q q \rangle$.
For $N_f>1$ and $SU(N_f)_A$ chiral symmetry is unbroken. So there should be ``molecules" $\bar L L $, 
similar to $\bar I I$ molecules originally discussed in \cite{Ilgenfritz:1988dh,Ilgenfritz:1994nt,Velkovsky:1997fe}, and
called ``bions" by Unsal and collaborators.

The opposite case is a dense ensemble at $T\sim T_c$, discussed by mean field
approximation in a number of settings \cite{Liu:2015ufa,Liu:2015jsa,Liu:2018znq}.
Here we will only discuss the numerical simulations for the $SU(2)$ 
in~\cite{Larsen:2015vaa,Larsen:2015tso}
and $SU(3)$ in~\cite{DeMartini:2021dfi,DeMartini:2021xkg}  color groups, without
($N_f=0$) and with two quark flavors ($N_f=2$) . 

\begin{figure}[htbp]
\begin{center}
\includegraphics[width=7cm]{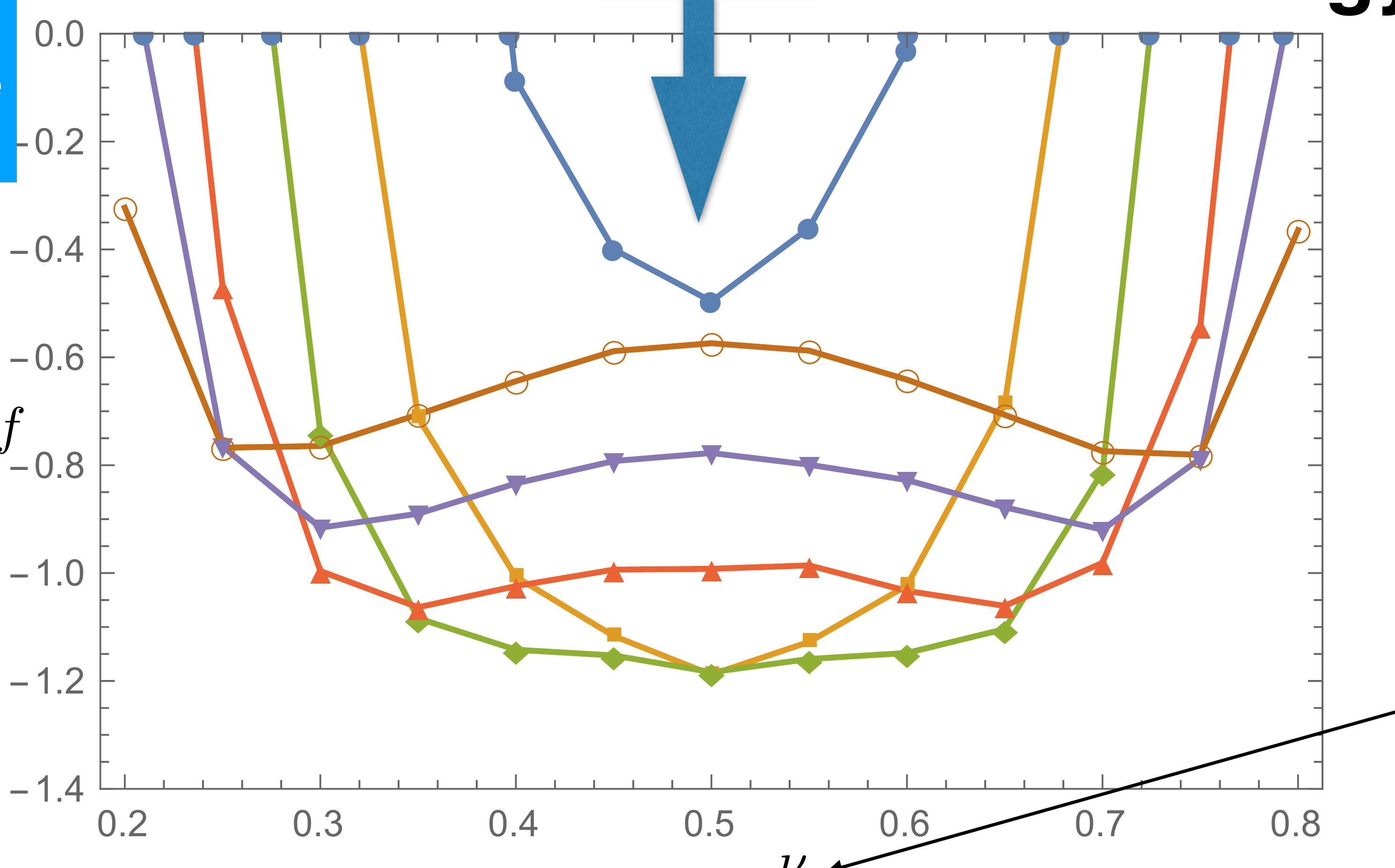}
\includegraphics[width=7cm]{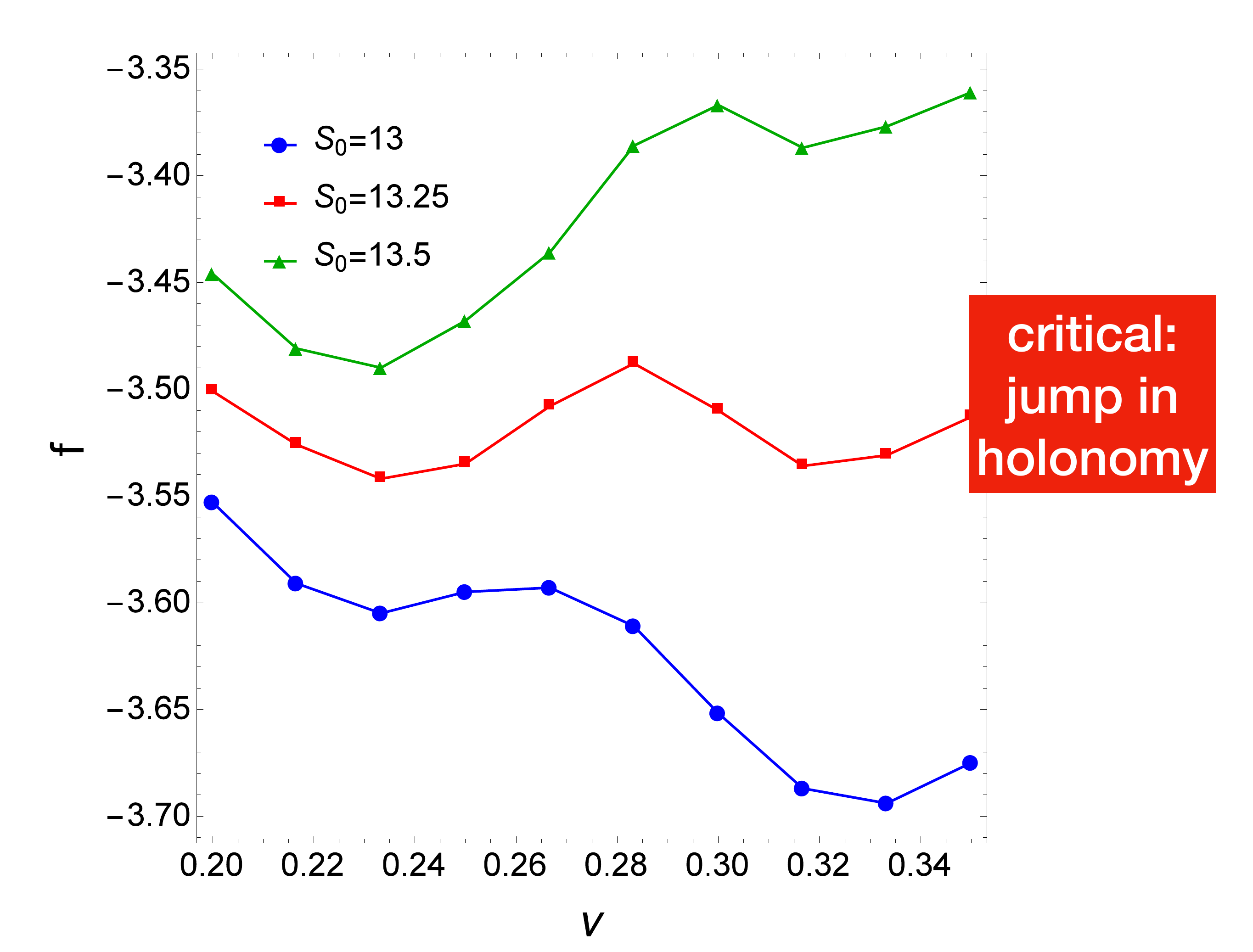}
\caption{Free energy $f$ versus the holonomy parameter $\nu$, for SU(2)
(top) and SU(3) (bottom) pure gauge theories. Different curves are for different
instanton densities (or temperatures). See text.
}
\label{fig_f_of_nu}
\end{center}
\end{figure}

The first physics issue is the $deconfinement$ phase transition. Recall that the Gross-Pisarski-Yaffe (GPY) perturbative potential for the Polyakov line, 
favors a trivial $\langle P \rangle =1$ case,  and disfavors confinement l $\langle P \rangle =0$. 
Therefore, to have confinement, the nonperturbative effects -- with a sufficient density of the dyons in the simulations -- should $overcome$ the GPY potential.
In Fig.\ref{fig_f_of_nu} we see that this happens differently for  SU(2)
 and SU(3) pure gauge theories. In the former the minimum gradually shifts to
the confining value  of $\nu=1/2$, and stay there at high densities. In the latter,
there is a jump, indicating first order transition. There is no place here
for comparison  with lattice data. We just mention  that the  SU(3) case $\langle P \rangle$ jumps to a value 0.4, 
remarkably close to the reported lattice value. The 
$deformation$ of the SU(3) gauge theory by an operator $\sim P^2$ in the instanton-dyon ensembles, is also 
in agreement with the lattice simulation.
In particular, the deconfinement temperature is observed to increase, with the size of the  jump  decreasing.

The second nonperturbative issue is chiral symmetry breaking at low $T$.
Again, it requires a sufficient density of the dyons, so that their zero modes can
get $collectivized$\footnote{Note that already in 1961,  the NJL model has shown that
one needs a large enough 4-fermion coupling to break chiral symmetry.}.

\begin{figure}[htbp]
\begin{center}
\includegraphics[width=15cm]{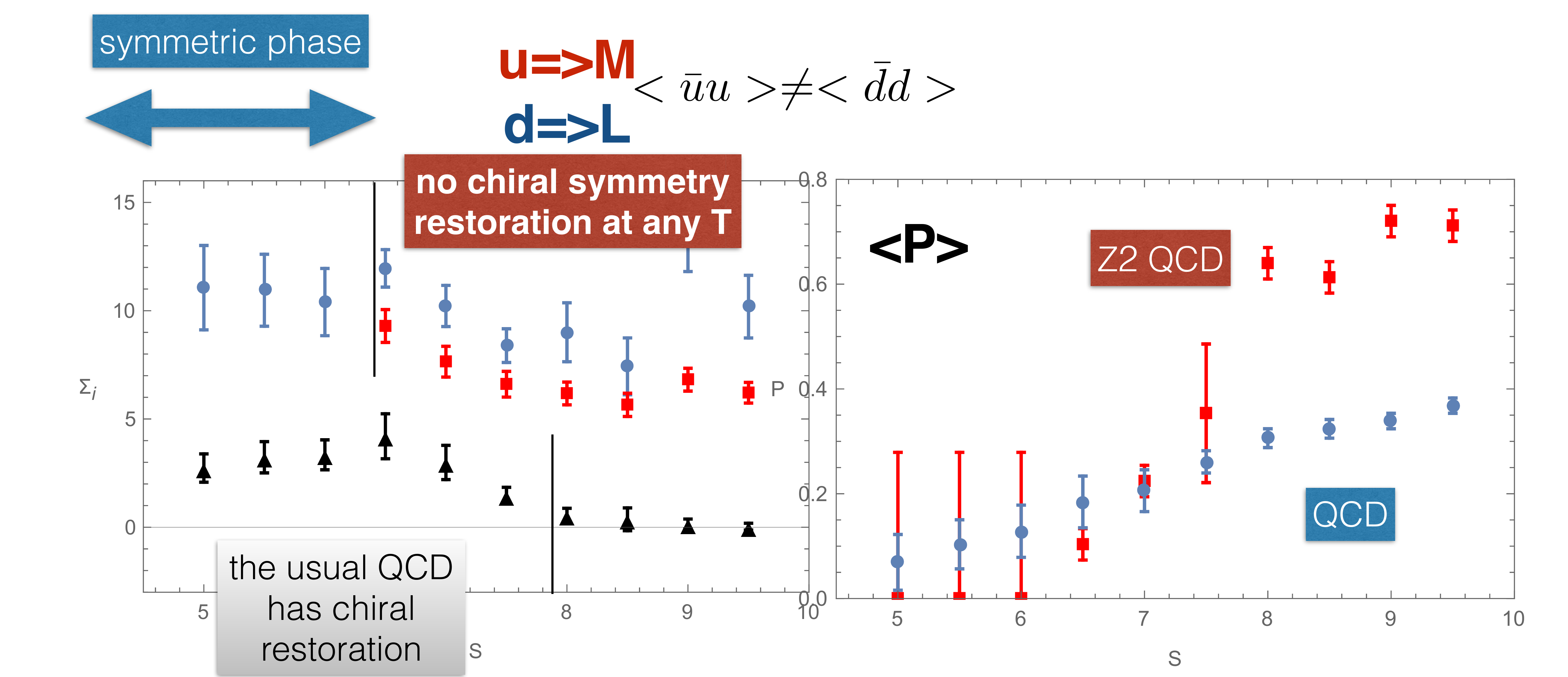}
\caption{Quark condensates (left) and Polyakov line (right)  
versus the action parameter $S(T)$. The  higher temperatures are recorded on the right vertical axis.
 The blue circles in the right plot are for QCD with  $N_c=N_f=2$, and the red squares are for $Z_2 QC_2D$.
 A $crossover$ in the former,  changes to a first order transition
 in the latter.
The black triangles on the left are for QCD with $N_c=N_f=2$,  with restored chiral $SU(2)_A$  symmetry above $S>8$. The blue discs and red squares 
show two chiral condensates for   $Z_2 QC_2D$. We clearly see a transition from a 
symmetric to an asymmetric phase, with no  chiral symmetry restoration. 
}
\label{fig_deformed}
\end{center}
\end{figure}

\subsection{QCD deformation by modified quark periodicity phases}
This deformation came first under the name of "imaginary chemical potentials"
 \cite{Roberge:1986mm}.  Its usage is different for $small$ and $large$ deformations.
In the former case, 
the motivation was due to the fact that imaginary chemical potentials 
can be simulated by usual Monte Carlo algorithms, while those with real $\mu$ cannot. 
A plot of the  lattice results for negative $\mu^2<0$, allows  the extrapolation to real $\mu$, e.g. by Taylor series.
This strategy has been used in many lattice studies.

%
%

 For a large phase,  Roberge and Weiss predicted a  first order transitions 
close to $z = (2k + 1)\pi/Nc$,  due to different 
  $N_c$ branches of the gluonic GPY potential. Of course, it is a perturbative
  argument expected to be true at large $T$ only. And indeed, when dyons are numerous this transition ends,
%
 and according to~\cite{Bonati:2016pwz} it happens at
$ T_{RW}=1.34 (7) T_c=  208(5)\,\,\, MeV $.

Another form of deformed QCD is to select imaginary chemical potentials  proportional to $T$, so that
the quark fugacities are  $exp(iz_f),f=1..N_f$,  with certain $T$-independent {\em periodic phases}. Moving
from conventional $z_f=\pi$ (quarks are fermions) to other values,  one should see multiple phase transitions,
each time when one of the $z_f$ crosses the Polyakov phases $\mu_i(T)$, as the fermion zero modes jump from
one instanton-dyon to another. The `ultimate" selection for $N_c=N_f$ theories was proposed in
  \cite{Kouno:2013zr}, with $z_f$ suggested to be located ``democratically", 
with one phase located  inside each of the holonomy
sectors $[\mu_{i+1},\mu_i]$  (for $T<T_{deconfinement}$).

 In Fig.\ref{fig_deformed} we show ensembles of instanton-dyons simulations for $N_c=N_f=2$
QCD and $Z_2QC_2D$,  with one quark being a fermion and one being a boson. The plots
show drastic changes in  both phase transitions. Deconfinement changes from
a crossover to a strong first order, and chiral restoration is not seen at all
\footnote{Note that  SU(2) is  somewhat of a special case. If the group is  SU(3), and two sectors 
of $M_1,M_2$ dyons would have $z_1,z_2$ located in them, ultimately  at high $T$ these sectors shrink as $\mu_i(T)$ move towards zero. When they cross
phases $z_1,z_2$ , the zero modes would return to $L$ dyons,  and thus at infinite $T$
 would become similar to undeformed QCD, with chiral symmetry  is restored.}. We note that 
the multiply deformed worlds (such as
 $Z_{N_c}QCD$) are unphysical, and qualitatively different from ours. We expect them to be separated by singularities.  Indeed, 
 we recall that  $Z_{N_c}QCD$ has completely different 
  flavor and chiral symmetries  (except at high $T$ when all $\mu_i$ are near zero). Its chiral symmetry is split diagonally for all flavors 
  $(U(1)_A)^{N_f-1}$, each broken 
$explicitly$ by diquark $^\prime$t Hooft operators for each dyon type. Obviously, this chiral symmetry breaking has no
relation to the {\em spontaneously broken $SU(N_f)$ chiral symmetry} occuring in our world, and  due to 
the multi-quark operators with  at finite instanton-dyon density. 

Another take on this unusual  phases with modified
quark periodicities,   historically came from supersymmetry.  Davies, Hollowood and Khose~\cite{Davies:1999uw} considered $\cal N$=1 SYM theory
on $R^{3}\times  S^1$ with a small circle and $bosonic$ gluinos, and evaluated the 
quark condensate using instanton-dyons. 
In this case the gluino term in the GPY potential changes sign, canceling the gluon contribution.

 Unsal~\cite{Unsal:2007jx} considered theories with {\em  more than one flavor} of gluinos (adjoint fermions),
$N_a>1$. The GPY potential  was found to be multiplied by $(1-N_a)$, and for  $N_a>1$ was $inverted$,
favoring confinement at weak coupling (small circle or high $T$). 
It was suggested that with confinement present both at small and large $L=1/T$, there would be $continuity$
(no phase transitions) at any $L$. However, lattice studies~\cite{Cossu:2009sq} have found $two$ deconfined phases in between
those two limits, preventing in general such continuity. The  considerations based on the GPY
potentials at weak coupling (high temperature),  do not necessarily carry to strong coupling.

\section{Summary}
The physics of instantons is now nearly 50 years old, yet we are still finding
 new applications in the vast  land of hadronic physics. Indeed, only 
recently, the large contributions due to the instanton-antiinstanton molecules came to  focus.
Recently, we used these contributions  to solve two old puzzles. Specifically,
the nonperturbative nature of the quark-antiquark
potentials,  and the  large observed mesonic formfactors in the semi-hard regime with  $Q^2\sim $ few $GeV^2$,
can be explained by these  fields in the QCD vacuum. 

The physics of diquarks, first appearing in connection to color superconductivity, again came
to the front of spectroscopy, with recent discovery of the double charmed
tetraquark state $T_{cc}$ made of $cc \bar u \bar d$.

The
development of light front Hamiltonians and wave functions entered a stage at which
properties of $all$ hadrons, from bottomonium to pions and with all kind of baryons and exotica
be systematically calculated.  Traditional distinction between heavy and light physics (absence
of nonrelativism in the former case) is no longer present on the light front, so all
of them are studied in the same setting.

Methods for solving the corresponding equations include
variational ones, diagonalization of Hamiltonian written as matrices in appropriate basis,
and even direct numerical solutions. This new "spectroscopy on the light front" program
 is going to bridge the gap between the usual spectroscopy, the
Euclidean version  (lattice and instantons) and the  light front observables. 

 The semiclassical theory based on ensembles of instanton-dyons,  reproduces semi-quantitatively
 the main lattice findings about deconfinement in pure gauge theories, and chiral phase transitions, in 
 QCD with light quarks.  Deforming QCD by extra action with powers of the Polyakov loop  shift/modify  
 the deconfinement transition.  Deforming it via quark periodicity phases,  leads to phases with drastically different 
 deconfinement and  chiral transitions.   While all these phases have multiple
 unphysical properties, their existence shed light on  the mechanisms driving  the
 QCD phase transitions. Their key feature are  ``jumps" of the fermion zero modes,
 from one dyon type to another.
 
 The dyon zero modes from lattice QCD simulations preserve remarkably well their shapes
 on the lattice, even in the case of their strong overlaps. Millions of thermal gluons do not seem to 
 affect them.
 
Finally, we note that  instanton-dyon simulations are simple multiple integrals over the instanton-dyon collective variables. So it is relatively easy
 to get to say $N_d\sim 300$ of instanton-dyons. In contrast, the  lattice simulations which are of course based on first principles, 
their current cost  limit them to about  $N_d\sim O(10)$ instanton-dyons, as  e.g.  seen from our Fig.2. If the instanton-dyon ensembles
are important for QCD phase transitions as we suggest,  then their current values for $N_d$ are perhaps too small to get an accurate 
description of their phases.

There seems to be enough new land to cover for the next 50 years.


\begin{thebibliography}{99}

%
%
%
%
%
%
%
%
%
%
%
%
%

\bibitem{Shuryak:2021vnj}
E.~Shuryak,
Lect. Notes Phys. \textbf{977} (2021), 1-520
doi:10.1007/978-3-030-62990-8

\bibitem{Shuryak:1987tr}
E.~V.~Shuryak,
Nucl. Phys. B \textbf{302} (1988), 621-644
doi:10.1016/0550-3213(88)90191-5

\bibitem{Sauter:1931zz}
F.~Sauter,
Z. Phys. \textbf{69} (1931), 742-764
doi:10.1007/BF01339461

\bibitem{Polyakov:1976fu}
A.~M.~Polyakov,
Nucl. Phys. B \textbf{120} (1977), 429-458
doi:10.1016/0550-3213(77)90086-4

\bibitem{Vainshtein:1981wh}
A.~I.~Vainshtein, V.~I.~Zakharov, V.~A.~Novikov and M.~A.~Shifman,
Sov. Phys. Usp. \textbf{25} (1982), 195
doi:10.1070/PU1982v025n04ABEH004533

\bibitem{Aleinikov:1987wx}
A.~A.~Aleinikov and E.~V.~Shuryak,
Yad. Fiz. \textbf{46} (1987), 122-129

\bibitem{Wohler:1994pg}
C.~F.~Wohler and E.~V.~Shuryak,
Phys. Lett. B \textbf{333} (1994), 467-470
doi:10.1016/0370-2693(94)90169-4
[arXiv:hep-ph/9402287 [hep-ph]].

\bibitem{Escobar-Ruiz:2015nsa}
M.~A.~Escobar-Ruiz, E.~Shuryak and A.~V.~Turbiner,
Phys. Rev. D \textbf{92} (2015) no.2, 025046
[erratum: Phys. Rev. D \textbf{92} (2015) no.8, 089902]
doi:10.1103/PhysRevD.92.025046
[arXiv:1501.03993 [hep-th]].

\bibitem{Belavin:1975fg}
A.~A.~Belavin, A.~M.~Polyakov, A.~S.~Schwartz and Y.~S.~Tyupkin,
Phys. Lett. B \textbf{59} (1975), 85-87
doi:10.1016/0370-2693(75)90163-X

\bibitem{tHooft:1976snw}
G.~'t Hooft,
Phys. Rev. D \textbf{14} (1976), 3432-3450
[erratum: Phys. Rev. D \textbf{18} (1978), 2199]
doi:10.1103/PhysRevD.14.3432

\bibitem{Shuryak:2019ikv}
E.~Shuryak and J.~M.~Torres-Rincon,
Phys. Rev. C \textbf{101} (2020) no.3, 034914
doi:10.1103/PhysRevC.101.034914
[arXiv:1910.08119 [nucl-th]].

\bibitem{Escobar-Ruiz:2016aqv}
M.~A.~Escobar-Ruiz, E.~Shuryak and A.~V.~Turbiner,
Phys. Rev. D \textbf{93} (2016) no.10, 105039
doi:10.1103/PhysRevD.93.105039
[arXiv:1601.03964 [hep-th]].

\bibitem{Nambu:1961tp}
Y.~Nambu and G.~Jona-Lasinio,
Phys. Rev. \textbf{122} (1961), 345-358
doi:10.1103/PhysRev.122.345

\bibitem{Shuryak:1982hk}
E.~V.~Shuryak,
Nucl. Phys. B \textbf{203} (1982), 140-156
doi:10.1016/0550-3213(82)90480-1

\bibitem{Schafer:1996wv}
T.~Sch\"afer and E.~V.~Shuryak,
Rev. Mod. Phys. \textbf{70} (1998), 323-426
doi:10.1103/RevModPhys.70.323
[arXiv:hep-ph/9610451 [hep-ph]].

\bibitem{Shuryak:1993kg}
E.~V.~Shuryak,
Rev. Mod. Phys. \textbf{65} (1993), 1-46
doi:10.1103/RevModPhys.65.1

\bibitem{Shuryak:1991pn}
E.~V.~Shuryak and J.~J.~M.~Verbaarschot,
Phys. Rev. Lett. \textbf{68} (1992), 2576-2579
doi:10.1103/PhysRevLett.68.2576

\bibitem{Shuryak:1992jz}
E.~V.~Shuryak and J.~J.~M.~Verbaarschot,
Nucl. Phys. B \textbf{410} (1993), 37-54
doi:10.1016/0550-3213(93)90572-7
[arXiv:hep-ph/9302238 [hep-ph]].

\bibitem{Shuryak:1992ke}
E.~V.~Shuryak and J.~J.~M.~Verbaarschot,
Nucl. Phys. B \textbf{410} (1993), 55-89
doi:10.1016/0550-3213(93)90573-8
[arXiv:hep-ph/9302239 [hep-ph]].

\bibitem{Schafer:1993ra}
T.~Sch\"afer, E.~V.~Shuryak and J.~J.~M.~Verbaarschot,
Nucl. Phys. B \textbf{412} (1994), 143-168
doi:10.1016/0550-3213(94)90497-9
[arXiv:hep-ph/9306220 [hep-ph]].


\bibitem{Thorsson:1989fw}
V.~Thorsson and I.~Zahed,
Phys. Rev. D \textbf{41}, 3442 (1990)
doi:10.1103/PhysRevD.41.3442



\bibitem{Schafer:2000et}
T.~Sch\"afer and E.~V.~Shuryak,
Lect. Notes Phys. \textbf{578} (2001), 203-217
[arXiv:nucl-th/0010049 [nucl-th]].

\bibitem{Ostrovsky:2002cg}
D.~M.~Ostrovsky, G.~W.~Carter and E.~V.~Shuryak,
Phys. Rev. D \textbf{66} (2002), 036004
doi:10.1103/PhysRevD.66.036004
[arXiv:hep-ph/0204224 [hep-ph]].

\bibitem{Shuryak:2021iqu}
E.~Shuryak and I.~Zahed,
[arXiv:2102.00256 [hep-ph]].

\bibitem{Verbaarschot:1991sq}
J.~J.~M.~Verbaarschot,
Nucl. Phys. B \textbf{362} (1991), 33-53
[erratum: Nucl. Phys. B \textbf{386} (1992), 236-236]
doi:10.1016/0550-3213(91)90554-B

\bibitem{Khoze:1991sa}
V.~V.~Khoze and A.~Ringwald,
CERN-TH-6082-91.

\bibitem{Shuryak:2002qz}
E.~Shuryak and I.~Zahed,
Phys. Rev. D \textbf{67} (2003), 014006
doi:10.1103/PhysRevD.67.014006
[arXiv:hep-ph/0206022 [hep-ph]].


\bibitem{Kock:2020frx}
A.~Kock, Y.~Liu and I.~Zahed,
Phys. Rev. D \textbf{102}, no.1, 014039 (2020)
doi:10.1103/PhysRevD.102.014039
[arXiv:2004.01595 [hep-ph]].

\bibitem{Kock:2021spt}
A.~Kock and I.~Zahed,
Phys. Rev. D \textbf{104}, no.11, 116028 (2021)
doi:10.1103/PhysRevD.104.116028
[arXiv:2110.06989 [hep-ph]].

\bibitem{Ji:2013dva}
X.~Ji,
Phys. Rev. Lett. \textbf{110}, 262002 (2013)
doi:10.1103/PhysRevLett.110.262002
[arXiv:1305.1539 [hep-ph]].

\bibitem{Liu:2021evw}
Y.~Liu and I.~Zahed,
[arXiv:2102.07248 [hep-ph]].



\bibitem{Shuryak:2019zhv}
E.~Shuryak,
Phys. Rev. D \textbf{100} (2019) no.11, 114018
doi:10.1103/PhysRevD.100.114018
[arXiv:1908.10270 [hep-ph]].

\bibitem{Dorokhov:1993fc}
A.~E.~Dorokhov and N.~I.~Kochelev,
Phys. Lett. B \textbf{304} (1993), 167-175
doi:10.1016/0370-2693(93)91417-L

\bibitem{Shuryak:2021hng}
E.~Shuryak and I.~Zahed,
[arXiv:2111.01775 [hep-ph]].

\bibitem{Ilgenfritz:1988dh}
E.~M.~Ilgenfritz and E.~V.~Shuryak,
Nucl. Phys. B \textbf{319} (1989), 511-520
doi:10.1016/0550-3213(89)90617-2

\bibitem{Rapp:1997zu}
R.~Rapp, T.~Sch\"afer, E.~V.~Shuryak and M.~Velkovsky,
Phys. Rev. Lett. \textbf{81} (1998), 53-56
doi:10.1103/PhysRevLett.81.53
[arXiv:hep-ph/9711396 [hep-ph]].

\bibitem{Athenodorou:2018jwu}
A.~Athenodorou, P.~Boucaud, F.~De Soto, J.~Rodr\'\i{}guez-Quintero and S.~Zafeiropoulos,
JHEP \textbf{02} (2018), 140
doi:10.1007/JHEP02(2018)140
[arXiv:1801.10155 [hep-lat]].

\bibitem{Shuryak:2020ktq}
E.~Shuryak and I.~Zahed,
Phys. Rev. D \textbf{103} (2021) no.5, 054028
doi:10.1103/PhysRevD.103.054028
[arXiv:2008.06169 [hep-ph]].

\bibitem{Callan:1978ye}
C.~G.~Callan, Jr., R.~F.~Dashen, D.~J.~Gross, F.~Wilczek and A.~Zee,
Phys. Rev. D \textbf{18} (1978), 4684
doi:10.1103/PhysRevD.18.4684

\bibitem{Eichten:1980mw}
E.~Eichten and F.~Feinberg,
Phys. Rev. D \textbf{23} (1981), 2724
doi:10.1103/PhysRevD.23.2724

\bibitem{Arvis:1983fp}
J.~F.~Arvis,
Phys. Lett. B \textbf{127} (1983), 106-108
doi:10.1016/0370-2693(83)91640-4

\bibitem{1508.02178}
T.~Kawanai and S.~Sasaki,
Phys. Rev. D \textbf{92} (2015) no.9, 094503
doi:10.1103/PhysRevD.92.094503
[arXiv:1508.02178 [hep-lat]].

\bibitem{Meggiolaro:1997mw}
E.~Meggiolaro,
Eur. Phys. J. C \textbf{4} (1998), 101-106
doi:10.1007/s100520050189
[arXiv:hep-th/9702186 [hep-th]].

\bibitem{Shuryak:2000df}
E.~V.~Shuryak and I.~Zahed,
Phys. Rev. D \textbf{62} (2000), 085014
doi:10.1103/PhysRevD.62.085014
[arXiv:hep-ph/0005152 [hep-ph]].

\bibitem{Giordano:2009su}
M.~Giordano and E.~Meggiolaro,
eCONF \textbf{C0906083} (2009), 31
[arXiv:0909.3710 [hep-ph]].

\bibitem{Shuryak:2021hng}
E.~Shuryak and I.~Zahed,
[arXiv:2111.01775 [hep-ph]].

\bibitem{Musakhanov:2021gof}
M.~Musakhanov and U.~Yakhshiev,
Int. J. Mod. Phys. E \textbf{30} (2021) no.11, 2141005
doi:10.1142/S0218301321410056
[arXiv:2103.16628 [hep-ph]].

\bibitem{Langfeld:2010nm}
K.~Langfeld and E.~M.~Ilgenfritz,
Nucl. Phys. B \textbf{848} (2011), 33-61
doi:10.1016/j.nuclphysb.2011.02.009
[arXiv:1012.1214 [hep-lat]].

\bibitem{Larsen:2018crg}
R.~N.~Larsen, S.~Sharma and E.~Shuryak,
Phys. Lett. B \textbf{794} (2019), 14-18
doi:10.1016/j.physletb.2019.05.019
[arXiv:1811.07914 [hep-lat]].

\bibitem{Larsen:2019sdi}
R.~N.~Larsen, S.~Sharma and E.~Shuryak,
Phys. Rev. D \textbf{102} (2020) no.3, 034501
doi:10.1103/PhysRevD.102.034501
[arXiv:1912.09141 [hep-lat]].

\bibitem{Ilgenfritz:1994nt}
E.~M.~Ilgenfritz and E.~V.~Shuryak,
Phys. Lett. B \textbf{325} (1994), 263-266
doi:10.1016/0370-2693(94)90007-8
[arXiv:hep-ph/9401285 [hep-ph]].

\bibitem{Velkovsky:1997fe}
M.~Velkovsky and E.~V.~Shuryak,
Phys. Lett. B \textbf{437} (1998), 398-402
doi:10.1016/S0370-2693(98)00930-7
[arXiv:hep-ph/9703345 [hep-ph]].

\bibitem{Liu:2015ufa}
Y.~Liu, E.~Shuryak and I.~Zahed,
Phys. Rev. D \textbf{92} (2015) no.8, 085006
doi:10.1103/PhysRevD.92.085006
[arXiv:1503.03058 [hep-ph]].

\bibitem{Liu:2015jsa}
Y.~Liu, E.~Shuryak and I.~Zahed,
Phys. Rev. D \textbf{92} (2015) no.8, 085007
doi:10.1103/PhysRevD.92.085007
[arXiv:1503.09148 [hep-ph]].

\bibitem{Liu:2018znq}
Y.~Liu, E.~Shuryak and I.~Zahed,
Phys. Rev. D \textbf{98} (2018) no.1, 014023
doi:10.1103/PhysRevD.98.014023
[arXiv:1802.00540 [hep-ph]].

\bibitem{Larsen:2015vaa}
R.~Larsen and E.~Shuryak,
Phys. Rev. D \textbf{92} (2015) no.9, 094022
doi:10.1103/PhysRevD.92.094022
[arXiv:1504.03341 [hep-ph]].

\bibitem{Larsen:2015tso}
R.~Larsen and E.~Shuryak,
Phys. Rev. D \textbf{93} (2016) no.5, 054029
doi:10.1103/PhysRevD.93.054029
[arXiv:1511.02237 [hep-ph]].

\bibitem{DeMartini:2021dfi}
D.~DeMartini and E.~Shuryak,
Phys. Rev. D \textbf{104} (2021) no.5, 054010
doi:10.1103/PhysRevD.104.054010
[arXiv:2102.11321 [hep-ph]].

\bibitem{DeMartini:2021xkg}
D.~DeMartini and E.~Shuryak,
Phys. Rev. D \textbf{104} (2021) no.9, 094031
doi:10.1103/PhysRevD.104.094031
[arXiv:2108.06353 [hep-ph]].

\bibitem{Roberge:1986mm}
A.~Roberge and N.~Weiss,
Nucl. Phys. B \textbf{275} (1986), 734-745
doi:10.1016/0550-3213(86)90582-1

\bibitem{Bonati:2016pwz}
C.~Bonati, M.~D'Elia, M.~Mariti, M.~Mesiti, F.~Negro and F.~Sanfilippo,
Phys. Rev. D \textbf{93} (2016) no.7, 074504
doi:10.1103/PhysRevD.93.074504
[arXiv:1602.01426 [hep-lat]].

\bibitem{Kouno:2013zr}
H.~Kouno, T.~Makiyama, T.~Sasaki, Y.~Sakai and M.~Yahiro,
J. Phys. G \textbf{40} (2013), 095003
doi:10.1088/0954-3899/40/9/095003
[arXiv:1301.4013 [hep-ph]].

\bibitem{Davies:1999uw}
N.~M.~Davies, T.~J.~Hollowood, V.~V.~Khoze and M.~P.~Mattis,
Nucl. Phys. B \textbf{559} (1999), 123-142
doi:10.1016/S0550-3213(99)00434-4
[arXiv:hep-th/9905015 [hep-th]].

\bibitem{Unsal:2007jx}
M.~Unsal,
Phys. Rev. D \textbf{80} (2009), 065001
doi:10.1103/PhysRevD.80.065001
[arXiv:0709.3269 [hep-th]].

\bibitem{Cossu:2009sq}
G.~Cossu and M.~D'Elia,
JHEP \textbf{07} (2009), 048
doi:10.1088/1126-6708/2009/07/048
[arXiv:0904.1353 [hep-lat]].

\bibitem{Cherman:2016hcd}
A.~Cherman, T.~Sch\"afer and M.~\"Unsal,
Phys. Rev. Lett. \textbf{117} (2016) no.8, 081601
doi:10.1103/PhysRevLett.117.081601
[arXiv:1604.06108 [hep-th]].

\end{thebibliography}

\end{document}